\documentclass[a4paper,12pt]{article}
\usepackage{amsmath,amssymb}
\usepackage{epsf}

\usepackage[
      colorlinks=true,
      linkcolor=red,
      urlcolor=blue,
      filecolor=red,
      linkcolor=blue,
      citecolor = red,
      pdfstartview=FitV,
      pdftitle={Brownian motion in AdS/CFT},
        pdfauthor={Jan de Boer, Veronika Hubeny, Mukund Rangamani, Masaki Shigemori},
        pdfsubject={Brownian motion},
        pdfkeywords={brownian motion, AdS/CFT},
        pdfpagemode=None,
        bookmarksopen=true,
    hypertex  
      ]{hyperref}
\usepackage{cite}
\usepackage{epsfig}

\makeatletter

\def\p{\partial}

\def\unit{{1\kern-.65ex {\rm l}}}
\def\1{{1\kern-.65ex {\rm l}}}
\def\ap{{\alpha'}}

\def\Re{\mathop{\mathrm{Re}}\nolimits}

\def\Tr{\mathop{\mathrm{Tr}}\nolimits}
\def\tr{\mathop{\mathrm{tr}}\nolimits}

\let\ev=\bracket
\let\Ev=\Bracket

\def\gammat{\gamma_s}
\def\Rt{R_s}
\def\kappat{\kappa_s}

\def\CD{{\cal D}}

\def\CG{{\cal G}}

\def\CN{{\cal N}}
\def\CO{{\cal O}}

\def\bbR{{\mathbb{R}}}

\newcount\hour \newcount\minute
\hour=\time \divide \hour by 60
\minute=\time
\def\now{%
\ifnum \hour<13
  \ifnum \hour=0 \advance \hour by 12 \number\hour:\else \number\hour:\fi%
     \ifnum \minute<10 0\fi%
     \number\minute%
\ A.M.%
\else \advance \hour by -12 \number\hour:%
  \ifnum \minute<10 0\fi%
  \number\minute%
  \ P.M.%
\fi%
}

\makeatother

\numberwithin{equation}{section}  

\title{\vspace{1.5cm}{\bf \Large Brownian motion in AdS/CFT}}
\date{December 30, 2008}

\author{
Jan de Boer$^a$, Veronika E. Hubeny$^b$, Mukund Rangamani$^b$,
Masaki Shigemori$^a$\\ \\
\small\sl $^a$ Institute for Theoretical Physics, University of Amsterdam\\[-1.5mm]
\small\sl Valckenierstraat 65, 1018 XE Amsterdam, The Netherlands\\
\small \sl $^b$  Centre for Particle Theory \& Department of
Mathematical Sciences,
\\[-1.5mm]
\small \sl Science Laboratories, South Road, Durham DH1 3LE, United Kingdom. \\ \\
 {\small{\tt J.deBoer@uva.nl}},
 {\small{\tt veronika.hubeny@durham.ac.uk}}, \\
 {\small{\tt mukund.rangamani@durham.ac.uk}},
{\small {\tt M.Shigemori@uva.nl}}
}

\begin{document}
\setlength{\baselineskip}{18pt}
\begin{titlepage}
\maketitle
\vspace{-1.5cm}
\begin{picture}(0,0)(0,0)
\put(350,300){ITFA-2008-51}
\put(350,285){DCPT-08/71}
\end{picture}
\vspace{.5cm}

\begin{abstract}
We study Brownian motion and the associated Langevin equation in
AdS/CFT\@.  The Brownian particle is realized in the bulk spacetime as a
probe fundamental string in an asymptotically AdS black hole background,
stretching between the AdS boundary and the horizon.  The modes on the
string are excited by the thermal black hole environment and
consequently the string endpoint at the boundary undergoes an erratic
motion, which is identified with an external quark in the boundary CFT
exhibiting Brownian motion. Semiclassically, the modes on the string are
thermally excited due to Hawking radiation, which translates into the
random force appearing in the boundary Langevin equation, while the
friction in the Langevin equation corresponds to the excitation on the
string being absorbed by the black hole.  We give a bulk proof of the
fluctuation-dissipation theorem relating the random force and friction.
This work can be regarded as a step toward understanding the quantum
microphysics underlying the fluid-gravity correspondence.
We also initiate a study of the properties of the effective
membrane or stretched horizon picture of black holes using our
bulk description of Brownian motion.

\end{abstract}
\thispagestyle{empty}
\setcounter{page}{0}
\end{titlepage}


\renewcommand{\thefootnote}{\arabic{footnote}}

\tableofcontents

\newpage
\section{Introduction}
\label{intro}

One of the interesting problems in statistical mechanics concerns
the understanding of the origin of macroscopic dissipation and the
approach to thermal equilibrium from microscopical point of view.
Conventionally, given a statistical system in the thermodynamic or
hydrodynamic limit, we imagine the collisions between the
microscopic constituents of our system as being responsible for
both of these macroscopic phenomena.  This kinetic theory based
picture is firmly anchored on the basic idea of Brownian
motion---in 1827, the botanist Robert Brown observed \cite{Brown}
under a microscope that tiny pollen particles suspended in water
undergo incessant irregular motion, which became known as the
Brownian motion.\footnote{Classic reviews on Brownian motion are
\cite{Uhlenbeck:1930zz, Chandrasekhar:1943ws, UhlenbeckII}.  For a
more complete list of references, see {\it e.g.}\
\cite{Dunkel:2008}.}  As is well-known now, this peculiar motion
is due to collisions with the fluid particles in random thermal
motion.  Therefore, any particle immersed in fluid at finite
temperature exhibits such Brownian motion, from a small pendulum
suspended in a dilute gas \cite{Kappler:1931} to a heavy particle
in quark-gluon plasma.  This universal phenomenon suggests that
the interaction with microscopic constituents is responsible for
dissipation and thermalization on macroscopic scales.

Since its advent, the holographic AdS/CFT correspondence
\cite{Maldacena:1997re, Gubser:1998bc, Witten:1998qj,
Aharony:1999ti} has been exploited to study the physics of
non-Abelian quark-gluon plasmas at finite temperature from bulk
gravitational physics, and {\it vice versa}.  The dual
gravitational description of strongly coupled gauge theories
provides an efficient way to study the thermodynamic properties
and the phase structure of the gauge theory. More recently, it has
become clear that one can also exploit the gravitational
description to understand the hydrodynamic regime of the
quark-gluon plasma, as was originally proposed in
\cite{Bhattacharyya:2008jc} and has been significantly developed
afterwards (see \cite{Son:2007vk} and references therein for
earlier work on hydrodynamics in the AdS/CFT context).  Namely,
the long-wavelength physics described by a hydrodynamical
Navier--Stokes equation on the boundary side is holographically
dual to the long-wavelength fluctuation of the horizons of
asymptotically AdS black hole spacetimes on the gravitational
side. This correspondence allows for a detailed quantitative study
of the plasma from the bulk, and {\it vice versa.}  It is thus a
natural question to ask whether one can obtain a holographic
description of Brownian motion, which is one step towards the
microphysics underlying thermodynamics and hydrodynamics.  The aim
of this paper is to answer this question in the affirmative.

One intrinsic reason to be interested in Brownian motion within a
holographic setting is to better understand the microscopic origin
of the thermodynamic properties of black holes. It has become
clear from the formulation of the AdS/CFT correspondence that one
has an in-principle solution to the problem of quantum dynamics of
black holes: we only need to solve the problem phrased in terms of
the dual field theory variables. However, it is fair to say that a
concrete quantitative understanding of the physics in these
contexts is still lacking. One of the most useful playgrounds for
understanding the quantum behavior of black holes has been the
arena of supersymmetric black holes \cite{Strominger:1996sh,
Maldacena:1997de}. Here we not only understand in many cases the
microscopic origin of black hole entropy, but also in a number of
cases have a bulk picture of the nature of the microscopic states
making up the black hole degeneracy.
In fact, from these various analyses, there emerges a rather intriguing
picture of a quantum black hole---the black hole microstates form a sort
of spacetime foam that replaces the region inside the horizon. Any
single microstate is horizon-free, but the typical microstates are
expected to exhibit the characteristic features of black hole
spacetimes, which has been confirmed explicitly for some concrete
systems; see \cite{Mathur:2005zp, Bena:2007kg, Skenderis:2008qn,
Balasubramanian:2008da} for reviews. Given this state of affairs one might probe
these microstates beyond equilibrium thermodynamics and ask how the
ensemble of them leads to dissipation and thermalization seen in a
thermal medium. Understanding the description of Brownian motion seems
then a natural step towards getting a handle on the problem.
%
%

Conversely, as mentioned above, the AdS/CFT correspondence has been
immensely useful in understanding many qualitative (and sometimes
quantitative) features of quark-gluon plasmas. The famous lower bound on
the ratio of shear-viscosity to entropy density for relativistic
hydrodynamic systems, $\eta/s \ge 1/4 \pi$ \cite{Kovtun:2004de} (see
also a review \cite{Son:2007vk}), has certainly played an important role
in obtaining a quantitative understanding of the dynamics of the
quark-gluon plasma produced at RHIC\@. Furthermore, studies of the
motion of quarks, mesons, and baryons in the quark-gluon plasma have
been carried out in the holographic framework starting with the seminal
papers \cite{Herzog:2006gh, Liu:2006ug, Gubser:2006bz, Herzog:2006se,
CasalderreySolana:2006rq, Gubser:2006nz, Liu:2006he,
CasalderreySolana:2007qw}, by considering the dynamics of probe strings
and D-branes in asymptotically AdS black hole spacetime---for a sample
of recent reviews on the subject, see \cite{RHICreviews}. The general
philosophy in these discussions was to use the probe dynamics to extract
the rates of energy loss and transverse momentum broadening in the
medium, which bear direct relevance to the physical problem of motion of
quarks and mesons in the quark-gluon plasma.
In such computations, the motion of an external quark in the quark-gluon
plasma is assumed to be described by a relativistic Langevin equation
\cite{Moore:2004tg}.  In the most basic form, the Langevin equation is
parametrized by two constants: the friction (drag force) coefficient
$\gamma$ and the magnitude of the random force
$\kappa$.\footnote{\label{ftnt:rel_Einst_rel} More precisely, in the
relativistic case, the random force has different magnitudes $\kappa_L$
and $\kappa_T$ in the directions transverse and longitudinal to the
momentum $p$. In the non-relativistic limit $p\to 0$, they are equal:
$\kappa_L=\kappa_T$.  The parameters $\gamma$ and $\kappa_L$ are related
to each other by the Einstein relation, under the assumption that the
Langevin dynamics holds and gives the J\"uttner distribution $e^{-\beta
E}$.  On the other hand, $\kappa_T$ is an independent parameter
\cite{Moore:2004tg, Gubser:2006nz}.} Furthermore, the random force is
assumed to be white noise.
By using the AdS/CFT realization of external quarks, Refs.\
\cite{Herzog:2006gh, Gubser:2006bz} determined the friction coefficient
$\gamma$, while Refs.\ \cite{CasalderreySolana:2006rq, Gubser:2006nz,
CasalderreySolana:2007qw} computed the random force
$\kappa$.\footnote{More precisely, \cite{Herzog:2006gh, Gubser:2006bz}
computed $\gamma$ and \cite{Gubser:2006nz, CasalderreySolana:2007qw}
computed $\kappa_T$, both in the relativistic case (the computation of
\cite{CasalderreySolana:2006rq} was nonrelativistic).  The longitudinal
component $\kappa_L$ does not have to be computed independently, since
it is related to $\gamma$ by the Einstein relation.  See also footnote
\ref{ftnt:rel_Einst_rel}.}

Therefore, one can say that the most basic data of the Langevin equation
describing Brownian motion in the CFT plasma are already available.
However, rather than taking such approaches which are phenomenological
in some sense, one could study more fundamental aspects of Brownian
motion in the AdS/CFT context.  For example, in the first place, why
does an external quark exhibit Brownian motion, and why is the motion
described by a Langevin equation?  
While the domain of validity of the Langevin equation is clear
from the previous results on the drag force, can we identify the origins
of the Brownian motion approximation from the bulk gravitational
description?  What is the bulk meaning of the fluctuation-dissipation
theorem relating $\gamma$ and $\kappa$?
The main purpose of the current paper is to elucidate the AdS/CFT
physics of Brownian motion, by addressing such questions.
For example, the computation of the random force in
\cite{CasalderreySolana:2006rq, Gubser:2006nz, CasalderreySolana:2007qw}
using the GKPW prescription \cite{Gubser:1998bc, Witten:1998qj} does not
explain what the bulk counterpart of the random force is.  We will see
that it corresponds to a version of Hawking radiation in the bulk.\footnote{For an earlier discussion of the relation between Hawking radiation and diffusion the context of AdS/QCD, 
see \cite{Myers:2007we}.} 

\begin{figure}[tbh]
\begin{quote}
 \begin{center}
 \epsfig{file=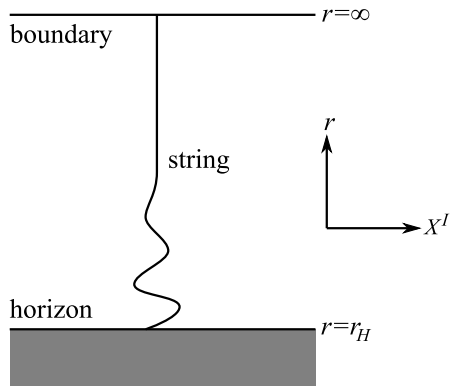,width=0.35\textwidth} \caption{\sl The
 bulk dual of a Brownian particle: a fundamental string hanging from the
 boundary of the AdS space and dipping into the horizon. The AdS black
 hole environment excites the modes on the string and, as a result,
 the string endpoint at infinity moves randomly, corresponding to the
 Brownian motion on the boundary. }
 \label{fig:brownianf1}
 \end{center}
\end{quote}
\end{figure}

Since we want to model Brownian motion, we need a gravitational analog
of a particle immersed in a thermal medium. In the boundary field theory
a natural particle is a test quark of large but finite mass immersed in
the quark-gluon plasma. This is realized in the dual gravitational
picture by introducing a fundamental string in the Schwarzschild-AdS
background. The endpoint of the string at the boundary then corresponds to
the test quark which undergoes Brownian motion; see Figure
\ref{fig:brownianf1}.


We will use this simple picture of a probe fundamental string in a
black hole background to ``derive'' the Brownian motion which the
string endpoint on the boundary undergoes.\footnote{A preliminary discussion of the fluctuations of a fundamental string in an asymptotically AdS black hole background can be found in \cite{Rey:1998bq}.} The basic idea is to
quantize the fluctuations of the string world-sheet about a
classical solution, which in the situations of interest
corresponds to a straight string hanging down from the boundary.
Since the bulk geometry has an event horizon, the induced metric
on the string world-sheet also corresponds to a black hole
geometry and the problem of studying fluctuations reduces to the
dynamics of two dimensional quantum fields in curved spacetime. By
quantizing the fluctuations we relate the quantum modes of the
string to the boundary endpoint. This mapping in principle allows
one to use the correlation functions of the position of the string
endpoint on the boundary to recover the excitation spectrum of the
string world-sheet. Assuming the validity of the semiclassical
approximation, we can then relate the thermal physics of the
Hawking radiation to the Brownian motion of the string endpoint
and derive the Langevin equation for the boundary dynamics.

The organization of the rest of the paper is as follows.  In
section \ref{sec:setup}, we set the stage for our discussion by
first reviewing the Langevin equation describing Brownian motion
in the field theory context. We then turn to a holographic
realization of Brownian motion in terms of the dynamics of a probe
fundamental string stretching from the boundary to the horizon of
an asymptotically AdS black hole geometry in $d$ dimensions.  We
write down the explicit relation between the boundary and bulk
quantities associated with the holographic Brownian motion.
This boundary-bulk relation can be explicitly worked out at the
semiclassical level, which we turn to in section
\ref{sec:semiclass}, focussing on  the simple case of three
dimensional spacetimes.  There, we assume that modes on the string
are thermally excited due to Hawking radiation and we derive the
Langevin dynamics exhibited by the boundary Brownian particle.
The friction and the random force appearing in the Langevin equation are
related to each other by the fluctuation-dissipation theorem.  In
section \ref{sec:FDthm}, we study this theorem from the bulk viewpoint
and give a bulk proof of it in the general case.
In section \ref{sec:gen_dim}, we generalize the discussion in
section \ref{sec:semiclass} for $d=3$ to general dimensions.
Despite being unable to quantize the modes on the string
analytically in this case we nevertheless show that at small
frequencies we recover the Langevin equation.
In section \ref{sec:BM_on_strtch_hor}, we study whether the bulk
Brownian motion of the fundamental string can be interpreted as
being caused by a suitable movement of the string endpoint on the
horizon, the idea being that the endpoint is randomly excited by
the stringy gas living on a membrane just outside the horizon,
much as in the spirit of the membrane paradigm. We also provide a
preliminary discussion of how to use our setup to study
microscopic properties of the stretched horizon. Ultimately, we
would like to directly probe properties of the quasi-particles that
make up the stringy gas living at the stretched horizon, but that
is beyond the scope of the present paper.
Section \ref{sec:disc} is devoted to a discussion.
Some of the relevant technical details are collected in the Appendices.

\section{Holographic Brownian motion}
\label{sec:setup}

To set the stage for our discussion we begin with a brief review
of the Langevin dynamics that describes the Brownian motion.  This
discussion will be the field theoretic, or boundary, side of the
story in the AdS/CFT context. Turning to the corresponding bulk
description, we will then describe how one can set up the problem
of studying the motion of a Brownian particle in a thermal medium
in terms of a probe string in an asymptotically Schwarzschild-AdS
black hole background.


\subsection{Brownian motion and Langevin dynamics}
\label{ss:BM_and_Lang}

Let us begin with the Langevin equation, which is the simplest model
describing a non-relativistic Brownian particle of mass $m$ in one
spatial dimension:
\begin{align}
 \dot p(t)&=-\gamma_0 \,p(t)+R(t),
\label{simpleLE}
\end{align}
where $p=m\dot x$ is the (non-relativistic) momentum of the
Brownian particle at position $x$, and $\dot{}\equiv d/dt$.  The
two terms on the right hand side of \eqref{simpleLE} correspond to
friction and a random force, respectively, and $\gamma_0$ is a
constant called the friction coefficient.  One can think of the
particle as losing energy to the medium due to the friction term
and concurrently getting a random kick from the thermal bath
modeled by the random force, which we assume to be white noise
with the following average:
\begin{align}
 \ev{R(t)}&=0,\qquad \ev{R(t)R(t')}=\kappa_0 \, \delta(t-t'),
 \label{RRcorrMarkovian}
\end{align}
where $\kappa_0$ is a constant.
The separation of the force into frictional and random parts on the
right hand side of \eqref{simpleLE} is merely a phenomenological
simplification---microscopically, the two forces have the same origin
(collision with the fluid constituents).

Assuming equipartition of energy, $\ev{m\, \dot x^2}=T$, with $T$ the
temperature,\footnote{We shall work in units where the Boltzmann
constant $k_B=1$.} one can derive the following time evolution for the
displacement squared \cite{Uhlenbeck:1930zz}:
\begin{align}
 \ev{s(t)^2}\equiv
 \ev{[x(t)-x(0)]^2}
 ={2\,D\over\gamma_0}\, \left(\gamma_0 \,t-1+e^{-\gamma_0 \,t}\right)
 \approx
 \begin{cases}
  \displaystyle {T\over m}\,t^2 & \displaystyle \Bigl(t\ll {1\over\gamma_0}\Bigr) \\[3ex]
  \displaystyle 2\,D\,t           & \displaystyle \Bigl(t\gg {1\over\gamma_0}\Bigr)
 \end{cases}
 \label{s^2_simple}
\end{align}
where the diffusion constant $D$ is related to the friction coefficient
$\gamma_0$ by the Sutherland--Einstein relation,
\begin{align}
 D&={T\over\gamma_0 \,m}.
 \label{diff_const_def}
\end{align}
We can see that in the ballistic regime, $t\ll 1/\gamma_0$, the particle moves
inertially ($s\sim t$) with the velocity determined by equipartition,
$\dot x\sim \sqrt{T/m}$, while in the diffusive regime, $t\gg
1/\gamma_0$, the particle undergoes a random walk ($s\sim\sqrt{t}$).  This is
because the Brownian particle must be hit by a certain number of fluid
particles to get substantially diverted from the direction of its initial velocity.
The crossover time between the two regimes is the relaxation time
\begin{align}
 t_{\text{relax}}&\sim {1\over\gamma_0},
\label{t_relax}
\end{align}
which characterizes the time scale for the Brownian particle to forget
its initial velocity and thermalize.
One can also derive the relation between the friction coefficient
$\gamma_0$ and the size of the random force $\kappa_0$
\begin{align}
 \gamma_0&={\kappa_0\over 2\,m\,T},
 \label{gamma_and_kappa}
\end{align}
which is the simplest example of the fluctuation-dissipation theorem and
arises due to the fact that the frictional and random forces are of the
same origin.

In $n$ spatial dimensions, $p$ and $R$ in \eqref{simpleLE} are
generalized to $n$ component vectors and \eqref{RRcorrMarkovian} is
generalized to
\begin{align}
 \ev{R_i(t)}&=0,\qquad \ev{R_i(t)R_j(t')}=\kappa_0\, \delta_{ij}\, \delta(t-t'),
\label{RRcorr_gendim}
\end{align}
where $i,j=1,\dots, n$.  In the diffusive regime, the displacement squared
goes as $\ev{s(t)^2}\approx 2nDt$.  The Sutherland--Einstein relation
\eqref{diff_const_def} and the fluctuation-dissipation relation \eqref{gamma_and_kappa} are
independent of $n$.

Now let us go back to the case with one spatial dimension ($n=1$).  The
Langevin equation \eqref{simpleLE}, \eqref{RRcorrMarkovian} captures
certain essential features of physics,
but nevertheless is too simple, for two reasons.  It assumes that the
friction is instantaneous and that there is no correlation between
random forces at different times (Eq.\ \eqref{RRcorrMarkovian}). If the
Brownian particle is not infinitely more massive than the fluid
particles, these assumptions are no longer valid; friction will depend
on the past history of the particle, and random forces at different
times will not be fully independent.  We can incorporate these effects
by generalizing the simplest Langevin equation \eqref{simpleLE} to the
so-called generalized Langevin equation \cite{Kubo:f-d_thm, Mori:genLE},
\begin{gather}
 \dot p(t)=-\int_{-\infty}^t dt'\, \gamma(t-t')\, p(t')+R(t)+K(t).
 \label{genLE}
\end{gather}
Now the friction term depends on the past trajectory via the memory
kernel $\gamma(t)$.  The random force is taken to satisfy
\begin{align}
 \ev{R(t)}=0,\qquad
 \ev{R(t)R(t')}=\kappa(t-t'),
 \label{RRcorr}
\end{align}
where $\kappa(t)$ is some function.  We have also now introduced an external
force $K(t)$ that can be applied to the system.

To analyze the physical content of the generalized Langevin equation we Fourier transform \eqref{genLE} to  obtain
\begin{align}
 p(\omega)={R(\omega)+K(\omega)\over \gamma[\omega]-i\omega},
 \label{genLE_omega}
\end{align}
where $p(\omega),R(\omega),K(\omega)$ are Fourier transforms, {\it
e.g.},
\begin{align}
 p(\omega)=\int_{-\infty}^\infty dt\, p(t)\,e^{i\omega t},
 \label{Fourier_trfm}
\end{align}
while
$\gamma[\omega]$ is the Fourier--Laplace transform:
\begin{align}
 \gamma[\omega]=\int_{0}^\infty dt\, \gamma(t) \,e^{i\omega t}.
 \label{FL_trfm}
\end{align}

If we take the statistical average of \eqref{genLE_omega}, the random force vanishes
because of the first equation in \eqref{RRcorr}, and we obtain
\begin{align}
 \ev{p(\omega)}=
 \mu(\omega)K(\omega),\qquad
 \mu(\omega)\equiv{1\over \gamma[\omega]-i\omega}.
\label{adm_def}
\end{align}
$\mu(\omega)$ is called the admittance.  So, we can determine the
admittance $\mu(\omega)$, and thereby $\gamma[\omega]$, by
measuring the response $\ev{p(\omega)}$ to an external force.  In
particular, if the external force is
\begin{align}
 K(t)=K_0 \,e^{-i\omega t},\label{monochr_force}
\end{align}
then $\ev{p(t)}$ is simply
\begin{align}
 \ev{p(t)}&=\mu(\omega)\, K_0\, e^{-i\omega t}.\label{presponse0}
\end{align}

For a quantity $\CO(t)$, define the power spectrum $I_\CO(\omega)$ by
\begin{align}
  I_\CO(\omega)=\int_{-\infty}^\infty dt\,\ev{\CO(t_0)\,\CO(t_0+t)}\,e^{i\omega t}.
 \label{pwrspctr_def}
\end{align}
Note that $\ev{\CO(t_0)\CO(t_0+t)}$ is independent of $t_0$ in a
stationary system.  The knowledge of power spectrum is the same as that
of 2-point function, because of the Wiener--Khintchine theorem
\begin{align}
 \ev{\CO(\omega)\CO(\omega')} =
 2\pi\delta(\omega+\omega')I_\CO(\omega).\label{WienerKhintchine}
\end{align}
Now consider the case without an external force, {\it i.e.},
$K=0$. In this case, from \eqref{genLE_omega},
\begin{align}
 p(\omega)={R(\omega)\over \gamma[\omega]-i\omega}.
\end{align}
Therefore, the power spectrum of $p$ and that for $R$ are related as
\begin{align}
 I_p(\omega)={I_R(\omega)\over |\gamma[\omega]-i\omega|^2}.\label{Ip_and_IR}
\end{align}

Combining \eqref{presponse0} and \eqref{Ip_and_IR}, one can determine
both $\gamma(t)$ and $\kappa(t)$ appearing in the Langevin equation
\eqref{genLE} and \eqref{RRcorr} separately.  However, as we will
discuss in section \ref{sec:FDthm}, these two quantities are not
independent but are related to each other by the fluctuation-dissipation
theorem, which is the generalization of the relation
\eqref{gamma_and_kappa}.

For the generalized Langevin equation, what corresponds to the
relaxation time \eqref{t_relax} is
\begin{align}
 {t_{\text{relax}}}&=\left[\int_0^\infty dt\,\gamma(t)\right]^{-1}
 ={1\over \gamma[\omega=0]}=\mu(\omega=0).
\label{t_relax_gen}
\end{align}
If $\gamma(t)$ is sharply peaked around $t=0$, we can ignore the
retarded effect of the friction term in \eqref{genLE} and write
\begin{align}
 \int_0^\infty dt'\,\gamma(t-t')\,p(t')
 \approx
 \int_0^\infty dt'\,\gamma(t')\cdot p(t)
 = {1 \over t_{\text{relax}}}\,p(t).
\end{align}
Then the Langevin equation reduces to the simple Langevin equation
\eqref{simpleLE} and it is clear that $t_{\text{relax}}$ corresponds to
the thermalization time for the Brownian particle.

Another physically relevant time scale, the microscopic (or collision
duration) time $t_{\text{coll}}$, is defined to be the width of the
random force correlator function $\kappa(t)$.  Specifically, let us
define
\begin{align}
 t_{\text{coll}} = \int_0^{\infty}\!\! dt \,{\kappa(t)\over \kappa(0)}.
 \label{def_t_coll}
\end{align}
If $\kappa(t)=\kappa(0)e^{-t/t_{\text{coll}}}$, the right hand side of
this precisely gives $t_{\text{coll}}$.  This $t_{\text{coll}}$
characterizes the time scale over which the random force is correlated,
and thus can be thought of as the time elapsed in a single process of
scattering.  In many cases,
\begin{align}
 t_{\text{relax}}\gg t_{\text{coll}}.
\label{tr>>tc}
\end{align}
Typical examples for which \eqref{tr>>tc} holds are the case where
the particle is scattered occasionally by dilute scatterers, and
the case where a heavy particle is hit frequently by much smaller
particles \cite{Kubo:f-d_thm}.  As we will see later, for the
Brownian motion dual to AdS black holes, the field theories are
strongly coupled CFTs and \eqref{tr>>tc} does not necessarily
hold.\

There is also a third natural time scale $t_{\text{mfp}}$ given by
the typical time elapsed between two collisions.  In the kinetic
theory, this mean free path time is typically $t_{\text{coll}} \ll
t_{\text{mfp}} \ll t_{\text{relax}}$; however in the case of
present interest, this separation no longer holds, as we will see.

\subsection{Bulk counterpart of Brownian motion}
\label{ss:Grav_Lang}

The AdS/CFT correspondence states that string theory in AdS$_{d}$
is dual to a CFT in $(d-1)$ dimensions.  In particular, the
(planar) Schwarzschild-AdS black hole with metric
\begin{align}
 ds_d^2&={r^2\over \ell^2}\left[-h(r)\,dt^2+d\vec X_{d-2}^2\right]
 +{\ell^2\over r^2 h(r)} \, dr^2,\qquad
 h(r)=1-\left({r_H\over r}\right)^{d-1}
 \label{AdSdBH}
\end{align}
is dual to a CFT at a temperature equal to the Hawking temperature
of the black hole,
\begin{align}
 T={1\over\beta}={(d-1)\,r_H\over 4\pi\,\ell^2}.
 \label{Hawk_temp_d}
\end{align}
In the above, $\ell$ is the AdS radius, and $t$,
$\vec{X}_{d-2}=(X^1,\dots, X^{d-2})\in \bbR^{d-2}$ are the boundary
coordinates.

In this black hole geometry \eqref{AdSdBH}, let us consider a
fundamental string suspended from the boundary at $r=\infty$,
straight down along the $r$ direction, into the horizon at
$r=r_H$; see Figure \ref{fig:brownianf1}.  In the boundary CFT,
this corresponds to having a very heavy external charged particle.
The $\vec X_{d-2}$ coordinates of the string at $r=\infty$ in the
bulk give the boundary position of the external particle.  As we
discussed above, such an external particle at finite temperature
$T$ is expected to undergo Brownian motion. The dual statement
must be that the black hole environment in the bulk excites the
modes on the string and, as the result, the endpoint of the string
at $r=\infty$ exhibits a Brownian motion which can be modeled by
a Langevin equation.

We study this motion of a string in the probe approximation where
we ignore its backreaction on the background geometry.  We assume
that there is no $B$-field in the background, which is the case
for AdS$_3$ based on D1- and D5-branes, and AdS$_5$ based on
D3-branes.  If we take the string coupling $g_s$ to be very small,
the interaction of the string with the thermal gas of closed
strings in the bulk of the AdS space can be ignored; the only
possible region with appreciable interaction is near the black
hole horizon which the string is dipping into.

Let us slightly generalize \eqref{AdSdBH} for a little while and
consider the following metric:
\begin{align}
 ds^2=g_{\mu\nu}(x)\, dx^\mu \, dx^\nu+G_{IJ}(x) \, dX^I \, dX^J,\label{genmetric}
\end{align}
where $x^\mu=t,r$ and $I,J=1,\dots,d-2$.  For the spacetimes of
interest, both $g_{\mu\nu}$ and $G_{IJ}$ are independent of
$X^I$.\footnote{We will mainly focus on planar black holes in AdS
corresponding to thermal field theories on $\bbR^{d-1,1}$ when the
transverse directions to the string $X^I$ are indeed Killing directions
in the bulk.}  Now, we stretch a string along the $r$ direction and
consider small fluctuation of it in the transverse directions $X^I$.
The action for the string is simply the Nambu--Goto action in the
absence of $B$-field.  In the gauge where the world-sheet coordinates
are identified with the spacetime coordinates $x^\mu=t,r$, the
transverse fluctuations $X^I$ become functions of $x^\mu$: $X^I=X^I(x)$.
If we expand the Nambu--Goto action up to quadratic order in $X^I$, we
obtain
\begin{align}
 S_{\text{NG}}&=
 -{1\over2\pi\ap}\int d^2x\sqrt{-\det\gamma_{\mu\nu}}
 \notag\\
 &\approx
 -{1\over 4\pi\ap}\int d^2x\sqrt{-g(x)}\,
 g^{\mu\nu}(x) \,  G_{IJ}(x) \,
 {\partial X^I\over \partial x^\mu} \,  {\partial X^J\over \partial x^\nu}
 \equiv S_{\text{NG}}^{(2)},
\label{SNG}
\end{align}
where $\gamma_{\mu\nu}$ is the induced metric, $g^{\mu\nu}$ is the
inverse of $g_{\mu\nu}$, and $g=\det g_{\mu\nu}$.  In the last line we
dropped the constant term that does not depend on $X^I$.  The quadratic
approximation is of course valid as long as the scalars $X^I$ do not
fluctuate too far from their equilibrium value (taken here to be $X^I
=0$).\footnote{One can show that when the modes on the string are
thermally excited in a black hole background at temperature $T$, this
quadratic approximation is valid outside the black hole except for the region within
$\sqrt{\ap}$ away from the horizon.}
In fact, this quadratic fluctuation Lagrangian for the world-sheet
scalars \eqref{SNG} can be thought of as taking the non-relativistic
limit; the Nambu--Goto action is after all a non-polynomial action in
the velocities $\partial_t X^I$ and we are expanding in the regime
$|\partial_t X^I |\ll 1$.  Therefore, we expect (and will see) that the
dual Langevin dynamics on the boundary will also be a non-relativistic
one, which is precisely what we reviewed in the previous subsection.
For most of the paper, we will use this quadratic action
$S_{\text{NG}}^{(2)}$ to study the fluctuations of the
string.\footnote{In Appendix \ref{sec:t_mfp}, we will consider the next
leading terms (quartic terms) when we estimate the mean free path time
$t_{\rm mfp}$.} The equation of motion derived from \eqref{SNG} is
\begin{align}
 0=\nabla^\mu(G_{IJ}\, \partial_\mu X^I)
 ={1\over\sqrt{- g}}\,\partial_{\mu}(\sqrt{-g}\,g^{\mu\nu}\,G_{IJ}\,\partial_\nu X^J),
\label{eom_NG}
\end{align}
where $\nabla_\mu$ is the covariant derivative with respect to
$g_{\mu\nu}$.  Note that this is not the same as the Klein--Gordon
equation in the spacetime \eqref{genmetric}, which would involve not
just $\partial_\mu=\partial/\partial x^\mu$ but also
$\partial_I=\partial/\partial X^I$.

Returning to the AdS black hole metric \eqref{AdSdBH}, we focus
first on the AdS$_3$ ($d=3$) case for simplicity (we will discuss
AdS$_d$ with general $d$ in section \ref{sec:gen_dim}) and study
the motion of a string in the black hole background.  In this
case, the metric \eqref{AdSdBH} becomes the nonrotating BTZ black
hole:
\begin{align}
 ds^2=-{r^2-r_H^2\over \ell^2}\, dt^2+{r^2\over \ell^2}\, dX^2
 +{\ell^2\over r^2-r_H^2}\,dr^2.
 \label{nonrotBTZ}
\end{align}
For the usual BTZ black hole, $X$ is written as $X=\ell \phi$ where
$\phi\cong\phi+2\pi$, but here we are taking $X\in\bbR$.
The Hawking temperature \eqref{Hawk_temp_d} is, in this case,
\begin{align}
 T\equiv{1\over\beta} ={r_H\over 2\pi\, \ell^2}.
 \label{Hawk_temp}
\end{align}
In terms of the tortoise coordinate $r_*$,
the metric \eqref{nonrotBTZ} becomes
\begin{align}
 ds^2={r^2-r_H^2\over \ell^2}\, (-dt^2+dr_*^2)+{r^2\over\ell^2}\, dX^2,
 \quad\qquad
 r_*\equiv{\ell^2\over 2r_H}\ln\left({r-r_H\over r+r_H}\right).
\label{tortoise}
\end{align}
For the BTZ metric \eqref{nonrotBTZ}, the equation of motion
\eqref{eom_NG} becomes
\begin{align}
 \left[-\partial_t^2
 +{r^2-r_H^2\over \ell^4\,r^2}\, \partial_r\Bigl(r^2\,(r^2-r_H^2)\, \partial_r\Bigr)\right]
 X(t,r)=0.\label{phi_eom}
\end{align}
As usual, we proceed by expanding $X$ in modes. Let us set
\begin{align}
 X(t,r)=e^{-i\omega t} f_\omega(r).
\end{align}
Then the equation of motion \eqref{phi_eom} can be written as
\begin{align}
\left[
{\nu^2}
+
{\rho^2-1\over \rho^2}\, \partial_\rho\Bigl( \rho^2(\rho^2-1)\partial_\rho\Bigr)
 \right] f_\omega=0,\label{eom_h_x}
\end{align}
where we defined dimensionless quantities
\begin{align}
 \rho\equiv {r\over r_H},\qquad
 \nu\equiv{\ell^2\omega\over r_H}={\beta \omega\over 2\pi}.
 \label{nu_def}
\end{align}
One can see that the linearly independent solutions to \eqref{eom_h_x}
are given by
\begin{gather}
 f_\omega^{(\pm)}
 = {1\over 1\pm i\nu}{\rho\pm i\nu \over \rho}\left({\rho-1\over \rho+1}\right)^{\pm i\nu/2}
 = {1\over 1\pm i\nu}{\rho\pm i\nu \over \rho}e^{\pm i\omega r_*}.\label{fpm_def}
\end{gather}
The normalization in \eqref{fpm_def} was chosen so that, near the
horizon,
\begin{align}
 f_\omega^{(\pm)}\sim  e^{\pm i\omega r_*}
 \qquad (\rho\sim 1),
\label{fpm_asympt}
\end{align}
and hence the solutions are written naturally in terms of ingoing
(``$+$'' sign) and outgoing (``$-$'' sign) modes.

\subsection{Boundary conditions and cut-offs}
\label{ss:BdyCut}

Before proceeding with the analysis of the fluctuations of the scalar
field $X$ in the BTZ geometry, it is useful to understand the boundary
conditions we want to impose on the fields.  While we are actually
interested in the world-sheet theory of the probe string, it is clear
that we can use the usual AdS/CFT rules to understand the boundary
conditions; in the static gauge the induced metric on the string
world-sheet inherits the geometric characteristics of an asymptotically
AdS$_{2}$ spacetime.

Usually in Lorentzian AdS/CFT one chooses to use normalizable
boundary conditions \cite{Balasubramanian:1998sn} for the modes.
However, in the present case, that would correspond to a string
extending all the way to $\rho=\infty$, which would mean that the
mass of the external particle is infinite and there would be no
Brownian motion.  So, instead, we have to impose a UV
cut-off\,\footnote{We use the terms ``UV'' and ``IR'' with respect
to the boundary energy.  In this terminology, in the bulk, UV
means near the boundary and IR means near the horizon.}  near the
boundary to make the mass finite.  Specifically, we implement this
by means of a Neumann boundary condition $\partial_r X=0$ at the
cut-off surface\footnote{In the AdS/QCD context, one can think of
the cut-off being determined by the location of the flavour brane,
whose purpose again is to introduce dynamical (and therefore
finite mass) quarks into the field theory.}
\begin{align}
 \rho=\rho_c\gg 1, \qquad \text{or}\qquad
 r=r_c \equiv r_H \rho_c.
\end{align}

The relation between the UV cut-off $\rho=\rho_c$ and the mass $m$ of the
external particle is easily computed from the tension of the string:
\begin{align}
 m&= {1\over 2\pi\ap}\int_{r_H}^{r_c} dr\,\sqrt{g_{tt}\,g_{rr}}
 ={r_c-r_H\over 2\pi\ap}
 ={\ell^2\,(\rho_c-1)\over \ap\beta}
 \approx {\ell^2\,\rho_c\over \ap\beta}.
 \label{m_and_rhoc}
\end{align}

Setting
\begin{align}
 f_\omega (\rho)= A \Bigl[f^{(+)}_\omega(\rho) + B f^{(-)}_\omega(\rho)\Bigr],
 \label{A[fp+Bfm]}
\end{align}
with constants $A$ and $B$, we obtain, on implementing the Neumann
boundary condition $\partial_\rho f_\omega|_{\rho=\rho_c}=0$,
\begin{align}
 B
 &= {1-i\nu \over 1+i\nu}\,{1+i \rho_c\nu \over 1 - i \rho_c\nu}
 \left({\rho_c-1\over \rho_c+1}\right)^{i\nu}
 \equiv e^{i\theta_\omega}.
 \label{BoverA}
\end{align}
Note that this is a pure phase.  This in particular means that, in the
near-horizon region $r\sim r_H$, we have, because of \eqref{fpm_asympt},
\begin{align}
 X(t,r) = f_\omega \, e^{-i\omega t}
 \sim  e^{-i\omega(t-r_*)}+
  e^{i\theta_\omega}\, e^{-i\omega(t+r_*)}.
\end{align}
The first term is a mode which is outgoing at the horizon, while
the second term is a mode reflected at $\rho=\rho_c$ and falling
back into the horizon, with phase shift $e^{i\theta_\omega}$.  The
fact that the outgoing and ingoing modes have the same amplitude
means that the AdS black hole, which Hawking radiates, can be in
thermal equilibrium at temperature $T$ \cite{Hemming:2000as}.
%

\begin{figure}[htbp]
  \begin{quote}
 \begin{center}
  \epsfxsize=5cm \epsfbox{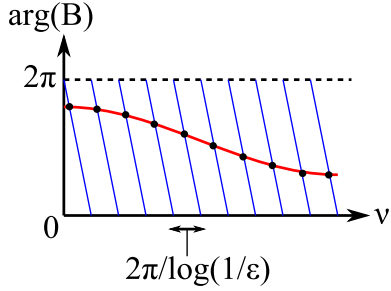} \caption{\sl Boundary conditions at
 infinity and horizon.  First, the UV boundary condition \eqref{BoverA}
 fixes $\arg B$ to lie, say, on the red line; at this point the possible
 values of $\nu$ are continuous. Further imposing the IR boundary
 condition \eqref{BoverAnh} makes the possible values of $\nu$ discrete
 (black dots).}  \label{fig:discrete}
\end{center}\end{quote}
\end{figure}

To regulate the theory, we need to introduce another cut-off near the
horizon $\rho=1$. Specifically, we cut off the geometry by putting an IR
cut-off (``stretched horizon'') at $\rho_s$
\begin{align}
 \rho_s=1+2\,\epsilon, \qquad \epsilon\ll1.
\end{align}
If we impose a Neumann boundary condition\footnote{One could also
take a Dirichlet boundary condition, but in the $\epsilon\to 0$
limit this would not make a difference.} at $\rho_s$, we have,
just as \eqref{BoverA},
\begin{align}
 B
 &= {1-i\nu \over 1+i\nu}\;{1+i(1+2\epsilon)\nu \over 1-i(1+2\epsilon)\nu}\,
 \epsilon^{i\nu}
 \approx \,  \epsilon^{i\nu}=e^{-i\nu\log(1/\epsilon)}.\label{BoverAnh}
\end{align}
If we require \eqref{BoverA} only, then we determine $B$ as a function
of $\nu$, but at this point $\nu$ can take any value
and is continuous.  If $\epsilon\ll 1$, further requiring
\eqref{BoverAnh} effectively makes the possible values of $\nu$
discrete, and the discreteness is given by
$\Delta \nu={2\pi/ \log(1/\epsilon)}\ll 1$;
see Figure \ref{fig:discrete}.
In terms of the frequency $\omega$, the discreteness is
\begin{align}
 \Delta \omega
 & ={4\pi^2 \over \beta\log(1/\epsilon)}.
\label{Deltaomega}
\end{align}
In other words, we have the following density of states:
\begin{align}
 \CD(\omega)={1\over \Delta \omega}
 &=
 {\beta\log(1/\epsilon) \over 4\pi^2}.\label{DoFomega}
\end{align}

All we have achieved by putting the regulator near the horizon is
to discretize the continuum spectrum which naturally occurs when
considering horizon dynamics.

Having done regularization, we can find a normalized basis of
modes and start quantizing $X(t,r)$ by expanding it in those
modes.  This is standard as in the case of a scalar field obeying
the Klein--Gordon equation; for details we refer the reader to
Appendix \ref{sec:norm_basis}\@. The upshot of the calculation is
\begin{align}
 X(t,r)
 &=
 \sum_{\omega>0}
 \left[a_\omega u_\omega(t,\rho) + a_\omega^\dagger u_\omega(t,\rho)^*\right],
 \label{X_expn}
\end{align}
where the summation is over $\omega$ discretized according to
\eqref{Deltaomega}.  The normalized basis $u_\omega$ is
\begin{align}
 u_\omega(t,\rho)=
 \sqrt{\ap\beta\over 2\,\ell^2\,\omega\, \log(1/\epsilon)}\,
 \Bigl[f^{(+)}_\omega(\rho)+B \, f^{(-)}_\omega(\rho)\Bigr] e^{-i\omega t},
\label{uomega_norm'd}
\end{align}
where $B$ is given by \eqref{BoverA}.  The expansion coefficients
$a_\omega$ satisfy the commutation relations
\begin{align}
 [a_\omega,a_{\omega'}]&= [a_\omega^\dagger,a_{\omega'}^\dagger]=0,\qquad
 [a_\omega,a_{\omega'}^\dagger]=\delta_{\omega\omega'}.
\end{align}

\subsection{The boundary-bulk dictionary}
\label{ss:bdybulk}

Given the behavior of quantum modes on the probe string in the
bulk, we can work out the dynamics of the endpoint, which
corresponds to a test quark in the thermal CFT plasma. To
understand the precise dictionary we look at the wave-functions of
the world-sheet fields ($X(t,\rho)$ in the BTZ geometry) in the
two interesting regions: (i) near the black hole horizon and (ii)
close to the boundary.

From \eqref{fpm_asympt}, near the horizon ($\rho\sim 1$), the expansion
\eqref{X_expn} becomes
\begin{align}
 X(t,\rho\sim 1)&\approx
\sum_{\omega>0}
 \sqrt{\ap\,\beta\over 2\,\ell^2\,\omega\,\log({1/\epsilon})}\,
 \Bigl[
 \left( e^{-i\omega(t-r_*)} + e^{i\theta_\omega}e^{-i\omega(t+r_*)} \right)\, a_\omega
 + {\rm h.c.}
 \Bigr]
 .\label{X_near_hor}
\end{align}
We see that the operators $a_\omega$ are directly related to the
amplitude for the outgoing modes, $e^{-i\omega(t-r_*)}$, near the
horizon.  On the other hand, at the UV cut-off $\rho=\rho_c$,
which we have chosen to be the location of the regularized
boundary, \eqref{X_expn} becomes
\begin{align}
 x(t)  &\equiv X(t,\rho_c)
 =\sum_{\omega>0}
 \sqrt{2\ap\,\beta\over  \ell^2\,\omega\,\log(1/\epsilon)}\,
 \left[
 {1-i\nu \over 1-i\rho_c \nu}\left({\rho_c-1\over \rho_c+1}\right)^{i\nu/2}\,e^{-i\omega t}\, a_\omega
 +{\rm h.c.}
 \right]
\label{X_rhoc}
\end{align}
This we will interpret as the position of the external particle (test
quark) in the boundary theory.  Here, the operators $a_\omega$ are
related to the Fourier coefficients of $x(t)$.

Using the above relation between \eqref{X_near_hor} and
\eqref{X_rhoc}, one can predict the correlators for the outgoing
modes near the horizon,
$\ev{a_{\omega_1}a_{\omega_2}^\dagger\dots}$ {\it etc.}, from the
boundary correlators $\ev{x(t_1)\, x(t_2)\dots}$ in field theory.
In particular, if we would be able to make a very precise
measurement of Brownian motion in field theory, we could \emph{in
principle} predict the precise state of the radiation that comes
out of the black hole.  In this way, we can learn about the
physics of quantum black holes in the bulk from the boundary data.
This of course requires us to compute the correlation function for
the test particle's position in a strongly coupled medium, which
is a difficult task that we will not undertake here.

However, at the semiclassical level, we can utilize this
dictionary to rather go from the bulk to the boundary and learn
about the boundary Brownian motion from the bulk data.
This is possible because, semiclassically, the state of the outgoing modes near the
horizon is given by the usual Hawking radiation.
As argued in \cite{Lawrence:1993sg, Frolov:2000kx}, the modes
on the string world-sheet which impinges on the black hole horizon are
thermally excited with a black-body spectrum determined by the Hawking
temperature. The quickest way to see this is to note that one can view
our analysis of the fluctuations of the string world-sheet \eqref{SNG}
as studying the dynamics of massless, free scalars in a two dimensional
black hole background.\footnote{The induced metric on the string
world-sheet clearly has a horizon and is an asymptotically AdS$_2$
spacetime.} Standard quantization of quantum fields in curved spacetime
\cite{Birrell:1982ix} will lead to the modes of the fields $X^I$ being
thermally excited at the Hawking temperature of this induced world-sheet
geometry which is the same as that for the BTZ black hole. In
particular, it follows that the outgoing mode correlators are determined
by the thermal density matrix
\begin{align}
 \rho_0&= {e^{-\beta H}\over \Tr(e^{-\beta H})},
 \qquad
 H=\sum_{\omega>0} \omega\, a_\omega^\dagger a_\omega.
 \label{rho0}
\end{align}
Note that, as we discussed above \eqref{genmetric}, here we are ignoring
the interaction of the string with the thermal gas of closed strings in
the bulk ($r_H<r<r_c$) of the black hole background.  Namely, we regard
the above density matrix \eqref{rho0} as solely due to the interaction
with the horizon.  
However, even if we took into account the weak interaction with the
thermal gas of closed strings, the density matrix would still be given
with very good accuracy by \eqref{rho0} because, at each value of $r$,
the thermal gas is in thermal equilibrium at the local Hawking
temperature and so is the string.

As long as we stay in the semi-classical approximation
we can
use the observations mentioned above to go from the bulk to the
boundary and derive the Brownian motion of the external particle
in the field theory. That is, instead of using the boundary field
theory to compute the correlation function of the quantum
operators $a$ and $a^\dagger$, we can use the fact that these
correlators are determined by the thermal physics of black holes
and utilize them to compute the boundary correlation functions. In
particular, we propose to use the knowledge \eqref{rho0} about the
outgoing mode correlators in the bulk, to predict the nature of
Brownian motion that the external particle on the boundary
undergoes. Thus by using the standard physics of black holes we
will be able to determine the functions $\gamma(t)$, $\kappa(t)$
appearing in the Langevin equation \eqref{genLE}.

\section{Hawking radiation and Brownian motion}
\label{sec:semiclass}

In the previous section, we used the AdS/CFT correspondence to
set-up a dictionary translating the information about the boundary
Brownian particle into corresponding data regarding the outgoing
modes (the world-sheet oscillators $a_I$ and $a_I^\dagger$ of the
fluctuations $X^I$) in the bulk.  In this section, we explicitly
derive the correlation function for the position of the test
particle. We \emph{assume} that the outgoing modes are the usual
Hawking radiation with the density matrix \eqref{rho0} and derive
the result that the endpoint at $\rho=\rho_c\gg 1$ indeed
undergoes a Brownian motion.

\subsection{Brownian motion of the boundary endpoint}
\label{sec:Brendpt}

Let us now consider the motion of the endpoint of the string at
$\rho=\rho_c\gg 1$. We will determine the behavior by computing
its displacement squared, which corresponds to \eqref{s^2_simple}.
In the canonical ensemble specified by  the density matrix
\eqref{rho0}, the relevant expectation values are given by the
Bose--Einstein distribution:
\begin{align}
 \ev{a_\omega^\dagger\, a_{\omega'}}
 =\Tr\left(\rho_0\, a_\omega^\dagger \,a_{\omega'}\right)
 ={\delta_{\omega\,\omega'}\over e^{\beta \omega}-1}.
\end{align}
Using this and \eqref{X_rhoc}, we compute
\begin{align}
  \ev{x(t)\,x(0)}
 =
  \ev{X(t,\rho_c)\, X(0,\rho_c)}
 &= {2\ap\,\beta\over \ell^2\,\log(1/\epsilon)}\; \sum_{\omega>0}
 \, {1\over\omega}\,{1+\nu^2\over 1+\rho_c^2\nu^2}
  \left({2\cos{\omega t}\over e^{\beta \omega}-1}+e^{-i\omega t}\right)\notag\\
 &= {\ap\,\beta^2 \over 2\pi^2 \,\ell^2}
 \, \int_0^\infty \,{d\omega\over\omega}\,
 {1+\nu^2\over 1+\rho_c^2\nu^2}\,
  \left({2\cos{\omega t}\over e^{\beta \omega}-1}+e^{-i\omega t}\right).
 \label{xx}
\end{align}
We are using the rescaled frequency $\nu$ defined in \eqref{nu_def}
throughout to avoid clutter.  In going to the second line, we utilized
the density of states determined in \eqref{Deltaomega} to rewrite the
sum as an integral.  From this, we compute the displacement of the
endpoint as:
\begin{align}
  s^2(t)
 &\equiv \ev{[x(t)-x(0)]^2}
 = {2\ap\beta^2 \over \pi^2 \ell^2}
 \int_0^\infty {d\omega\over\omega}\,
 {1+\nu^2\over 1+\rho_c^2\nu^2}
  \coth{\beta \omega\over 2}\,\sin^2{\omega t\over 2}.
 \label{Delta_phi^2}
\end{align}
This has a logarithmic UV divergence.  Because this divergence is
coming from the zero point energy (the $e^{-i\omega t}$ term in
\eqref{xx}), which exists even at zero temperature, we simply
regularize it by normal ordering the $a,a^\dagger$ oscillators
(${:\! a_\omega a_\omega^\dagger \!:} \equiv {:\! a_\omega^\dagger
a_\omega\! :}$).   When so regularized,
the correlator
\eqref{xx} becomes
\begin{align}
  \ev{{:\!x(t)x(0)\!:}}
 &= {\ap \beta^2 \over \pi^2 \,\ell^2}
 \int_0^\infty \,{d\omega\over\omega}\,
 {1+\nu^2 \over 1+\rho_c^2\nu^2}\;
  {\cos{\omega t}\over e^{\beta \omega}-1},
 \label{xx_reg}
\end{align}
and the displacement squared \eqref{Delta_phi^2} becomes\footnote{Note
that regularization by normal-ordering does not preserve the KMS
relations except in the classical limit.}$^,$\footnote{Another way to
regularize the correlator is to use the canonical correlator introduced
in \eqref{def_can_corr}.  Using \eqref{xx_can}, one can derive
\begin{align*}
 s^2_c(t)\equiv \ev{[x(t)-x(0)];[x(t)-x(0)]}
 &= {2\ap\beta^2 \over \pi^2 \,\ell^2}
 \int_{-\infty}^\infty\, {d\omega\over\omega^2}\,
 {1+\nu^2\over 1+\rho_c^2\nu^2}\;
  {\sin^2{\omega t\over 2}}\\
 &={\ap\beta^2\over \pi \ell^2}\left[
 (|t|/ \beta)-(1-{ \rho_c^{-2}})(1-e^{-{2\pi |t|/\beta \rho_c}})
 \right].
\end{align*}
This is finite and has exactly the same short- and long-time behaviors
as in \eqref{s_reg^2_behavior}.  Note, however, that the divergence
\eqref{Delta_phi^2} is related to the well-known fact that the
fluctuation of the position of a string always diverges
\cite{Karliner:1988hd}.  }
\begin{align}
  s^2_{\rm reg}(t)
 &\equiv
 \ev{{:\![x(t)-x(0)]^2\!:}}
 = {4\ap\,\beta^2 \over \pi^2 \,\ell^2}
 \int_0^\infty\, {d\omega\over\omega}\,
 {1+\nu^2\over 1+\rho_c^2\nu^2}\;
  {\sin^2{\omega t\over 2}\over e^{\beta \omega}-1}.
 \label{s_reg^2}
\end{align}
We analytically evaluate this integral in Appendix
\ref{sec:eval_s^2}\@. For the present purposes we will only record the
result for $\rho_c\gg 1$, which is all that is relevant for the
physics of the boundary field theory. We find the following behavior:
\begin{align}
 s^2_{\rm reg}(t) \approx
 \begin{cases}
  \displaystyle {\ap\over \ell^2\,\rho_c}\;t^2\approx {T\over m}\;t^2 \quad & (t\ll t_c), \\[2.5ex]
  \displaystyle {\ap\over \pi\, \ell^2\, T}\; t   \quad & (t\gg t_c).
\end{cases}
\label{s_reg^2_behavior}
\end{align}
So, we observe two regimes, the ballistic and diffusive regimes, exactly as
for the standard Brownian motion \eqref{s^2_simple}. The crossover
time $t_c$ is given by
\begin{align}
 t_c \sim  \beta\,\rho_c \,\sim {\ap\, m\over \ell^2\, T^2}.\label{crossovertime}
\end{align}

In the ballistic regime, $t\ll t_c$, the coefficient of $t^2$ in
\eqref{s_reg^2_behavior} is exactly the same as \eqref{s^2_simple}
determined by the equipartition of energy $\dot x\sim \sqrt{T/m}$.  In
fact, one can say much more; in Appendix \ref{sec:p-distribution}, we
show that, if $\rho_c\gg 1$, the probability distribution $f(p)$ for the
``momentum'' $p\equiv m\,\dot x$ of the endpoint is exactly equal to the
Maxwell--Boltzmann distribution for non-relativistic particles,
\begin{align}
 f(p)&\propto e^{-\beta E_p},\qquad
 E_p={p^2\over 2m}.
\end{align}

In the diffusive regime, $t\gg t_c$, we find a diffusion constant (one half of
the coefficient of $t$)
\begin{align}
 D_{\text{AdS$_3$}}= {\ap\over 2\pi\ell^2 T} \ .
 \label{diff_const_AdS3}
\end{align}
{\it A priori\/} this looks counterintuitive because it is inversely
proportional to temperature $T$ and implies that the random walk becomes
more vigorous at lower temperature.  However, this is consistent with
the known results for test quarks moving in the thermal $\CN =4$ super
Yang-Mills plasma \cite{Herzog:2006gh, Liu:2006ug, Gubser:2006bz,
CasalderreySolana:2006rq}.  For example, refs.\ \cite{Herzog:2006gh,
Gubser:2006bz} considered a heavy particle on the boundary moving at a
constant speed $v$ under the influence of an external force. One can
compute the friction acting on it from the bulk using the Nambu--Goto
action, where a string is moving at velocity $v$, trailing along the
boundary.  It is easy to generalize their computation to AdS$_d$ with
general $d$, the result being\footnote{Although this is a
straightforward generalization of \cite{Herzog:2006gh, Gubser:2006bz}
and the general formalism has been laid out in \cite{Herzog:2006se}, it
seems to us that this result has not appeared explicitly in the
literature.}
\begin{align}
 \dot p = - {8\pi \,\ell^2 \,T^2\over (d-1)^2\,\ap}
 \, {v\over (1-v^2)^{2/(d-1)}},\qquad
 p={m\,v\over \sqrt{1-v^2}}.
\end{align}
In the non-relativistic limit, $v\ll 1$, this means that the friction
constant is
\begin{align}
 \gamma_{0}^{\text{AdS$_d$}}=
 {8\pi \,\ell^2 \,T^2\over (d-1)^2\,\ap \,m}.\label{gamma_AdSd}
\end{align}
If we use the Sutherland--Einstein relation
\eqref{gamma_and_kappa},\footnote{Note that, as explained around
\eqref{RRcorr_gendim}, the relation \eqref{gamma_and_kappa} does not
depend on $d$.}  we obtain the diffusion constant
\begin{align}
 D_{{\rm AdS}_d}&={(d-1)^2 \,\ap\over 8\pi\, \ell^2 \, T},
 \label{diff_const_AdSd}
\end{align}
which agrees with \eqref{diff_const_AdS3} for $d=3$.

One can give an intuitive explanation for the reason why the diffusion
constant is inversely proportional to $T$ from the boundary viewpoint of
Brownian motion.  The random walk behavior of Brownian motion is due to
frequent collisions of the Brownian particle with the fluid particles.
In particular, after $n$ steps (collisions), the distance $s$ that a
random-walk particle covers scales as $\sqrt{n}\,L_{\rm mfp}$, where the
mean free path $L_{\rm mfp}$ is the typical length traveled between the
collisions, {\it i.e.}, it provides a scale for the system.  For the
thermal system under consideration we have, $L_{\rm mfp} \sim 1/T$,
because this is the only scale available in a CFT at temperature
$T$.\footnote{The precise value of the $L_{\rm mfp}$ depends on the
strength of the field theory coupling, but the temperature dependence
follows via dimensional analysis.  In fact, in Appendix \ref{sec:t_mfp},
we estimate the mean free path time to be $t_{\rm mfp}\sim
1/(\sqrt{\lambda}\,T)$, where $\lambda\sim \ell^4/\ap{}^2$.  If the plasma
constituents are moving at the speed of light, this means that $L_{\rm
mfp}\sim 1/(\sqrt{\lambda}\,T)$.  With this value of $L_{\rm mfp}$, we
can even recover the $\lambda$ dependence of \eqref{diff_const_AdSd}.}
After $n$ collisions, the time elapsed is given by $t \sim n/T$, since
the time between collisions is also given by $L_{\rm mfp} \sim 1/T$.
So, putting things together, we have $s \sim \sqrt{t T}\cdot
1/T=\sqrt{t/T}$, namely, $s^2 \sim t / T$ which is exactly what we infer
from \eqref{diff_const_AdSd}.

From the bulk point of view, on the other hand, one can give a physical
explanation for $D\sim 1/T$ as follows.  Near the horizon ($\rho\sim
1$), the Nambu--Goto action \eqref{SNG} becomes
\begin{align}
 S_{\text{NG}} \approx {\pi \,\ell^2\, T^2\over \ap}\,\int dt\,dr_* 
 \left[ (\partial_t X)^2 -(\partial_{r_*} X)^2 \right].\label{SNG_NH}
\end{align}
This is the same as the action for a string in flat space, with
$\ap$ replaced by $\alpha'_{\rm eff}=\ap/(4\pi^2\ell^2T^2)$.  This
means that the size of the fluctuations in $X^2$ is proportional
to $\alpha'_{\rm eff}\sim T^{-2}$.  The Boltzmann factor of
Hawking radiation gives an additional factor of $1/(e^{\beta\,
\omega}-1)$ which scales as $T$ at low frequency.  Altogether,
near the horizon, the fluctuations scale with temperature as
$X^2\sim T^{-1}$. When a fluctuation propagates to $\rho=\rho_c$,
a greybody factor damps the fluctuation.  However, as one can see
from \eqref{X_rhoc}, the damping is $\CO(1)$ for very small
frequency. This leads to $x^2=X(\rho=\rho_c)^2$ being $\sim
T^{-1}$. The reason why very low frequency modes can reach
$\rho=\rho_c$ undamped is that $X$ is an isometry direction and
very low frequency $X$ modes can propagate at almost no cost in
energy.

A natural question to ask is what happens to this $T^{-1}$ scaling
as $T\to 0$, as we expect that the endpoint should not fluctuate
at $T=0$. This can be understood by realizing that for a mode to
propagate to $\rho=\rho_c$ undamped, it should be the lightest
mode in the problem. In particular, only modes whose frequencies
are lower than the thermal scale (which goes to zero as $T\to 0$)
can propagate without damping.  Translating to real time dynamics,
this means that one needs to wait until $t\sim t_c$ to see the
diffusive regime; but since $t_c\propto T^{-2}\to \infty$ as $T\to
0$ we never enter that regime and the motion is always ballistic
as expected.

Thus, we have demonstrated that the endpoint of the string at
$\rho=\rho_c$ indeed behaves like a Brownian particle; it shows
ballistic and diffusive regimes, just as for the usual Brownian
motion.  We would now like to understand the Langevin equation
from the bulk perspective. As we will see below, the Langevin
equation governing this Brownian motion turns out to be not of the
simplest type \eqref{simpleLE} and \eqref{RRcorrMarkovian}, but
rather the generalized one \eqref{genLE} indicating that the
precise nature of the random kick encountered by the Brownian
particle depends on the past history of its trajectory.

\subsection{Forced motion \& the holographic Langevin equation}
\label{sec:Forcedm}

As we discussed in subsection \ref{ss:BM_and_Lang}, the
generalized Langevin equation \eqref{genLE} has two functional
parameters: the memory kernel $\gamma(t)$ and the auto-correlation
function $\kappa(t)$, related to the dissipative and stochastic
components, respectively.  We would like to determine these
functions from the holographic viewpoint for the probe string in
the black hole background. In order to do so, we will first
determine $\gamma(t)$, or equivalently $\mu(\omega)$.  Once we
know $\mu(\omega)$, we can compute $\kappa(\omega)$ by using
equation \eqref{Ip_and_IR} and the $\ev{x\,x}$ correlator
\eqref{xx} (or \eqref{xx_reg}).

We first turn to the determination of $\mu(\omega)$.  Consider
applying an external force on the Brownian particle as in
\eqref{monochr_force}; from the response to this force we can read
off $\mu(\omega)$ using \eqref{presponse0}.  So the natural
question is what external force is to be applied to the string
endpoint. As in the AdS/QCD set-ups, we can realize such forced
motion by placing a ``flavor'' D-brane at the UV cut-off
$\rho=\rho_c$ and by turning on world-volume electric field on it.
Since the endpoint of the string is charged, this will exert the
desired force on the Brownian particle.

So, let us consider the Nambu--Goto action \eqref{SNG} in the general
metric \eqref{genmetric}, and add to it
the following boundary term
\begin{align}
 S_{\text{bdy}}&=\oint A(x,X),
\end{align}
which corresponds to turning on world-volume field on the flavor D-brane (which is placed
at the UV cut-off $\rho_c$).  Here, $A(x,X)$ is a 1-form defined on the flavor
D-brane world-volume.  We again work in the gauge where the world-sheet coordinates are
identified with the spacetime coordinates $x^\mu=t,r$.  We have $X^I=X^I(x)$
and
\begin{align}
 S_{\text{bdy}}&
 =\oint [A_t(x,X)+A_I(x,X)\, \dot X^I] \, dt,\label{Sbndy}
\end{align}
where $t$ is taken to be the coordinate along the boundary as before (or
equivalently, the boundary is at $r=\text{const.}$).  The equation of
motion one obtains for the total action
$S_{\text{NG}}^{(2)}+S_{\text{bdy}}$ at the boundary is
\begin{align}
 \sqrt{-\widetilde g\,}\,n^\mu \,G_{IJ} \,  {\partial_\mu X^J}
 -  2\pi\ap\,\left(F_{It}+F_{IJ}\, \partial_t X^J\right)=0,
\label{gaon8Oct08}
\end{align}
where ${\widetilde g}_{\mu\nu}$ is the induced metric on the boundary, $n^\mu$ is the
outward-pointing unit normal to the boundary, and $F_{It}=\partial_I
A_t-\partial_t A_I$, $F_{IJ}=\partial_I A_J-\partial_J A_I$.

Returning to the simple setting of the BTZ geometry \eqref{nonrotBTZ},
the equation of motion for the string in the presence of this additional
gauge field is
\begin{align}
 \rho^2(\rho^2-1)\,\partial_\rho X
 = {2\pi\ap\,\ell^4\over r_H^3}\;F_{Xt}
 \qquad\qquad \text{at $\rho=\rho_c$}.
\label{bc_forced_x=xc}
\end{align}
For the world-volume field $F_{Xt}$ we choose an oscillating
electric field with frequency $\omega$:
\begin{align}
  F_{Xt}\equiv E=E_0 \, e^{-i\omega t},\label{monochr_wvelflx}
\end{align}
motivated by  \eqref{monochr_force}.  We now want to compute how
the string, in particular its endpoint $X(t,\rho_c)=x(t)$, moves
under the influence of this external force in order to compute the
admittance $\mu(\omega)$.

As before, the solution to the bulk equation of motion can be written as a linear combination of the modes $f^{(\pm)}_\omega(\rho)$:
\begin{align}
 X(t,\rho)
 &=
 \left[A' f^{(+)}_\omega(\rho) + B' f^{(-)}_\omega(\rho)\right]e^{-i\omega t}.
\label{X(t,r)_forced}
\end{align}
To determine the coefficients $A',B'$, we need to impose a boundary
condition at the horizon, in addition to the boundary condition
\eqref{bc_forced_x=xc} at $\rho=\rho_c$.  Effectively all that the
external field has done was to modify the Neumann boundary condition, which
we imposed earlier on the cut-off surface, to a mixed boundary
condition.

In the semiclassical approximation, the boundary condition near
the horizon is such that outgoing modes are always thermally
excited because of Hawking radiation, while the ingoing modes can
be arbitrary.\footnote{If one ignores the thermal excitations of
the outgoing modes and sets them to zero, this boundary condition
becomes the so-called purely ingoing boundary condition.}  From
\eqref{fpm_asympt}, the coefficients $A'$ and $B'$ correspond to
outgoing and ingoing modes respectively.  Therefore, the boundary
condition at the horizon is that $A'$ is thermally excited.
However, because the radiation is random, the phase of $A'$ takes
random values and, on average, $A'$ vanishes: $\ev{A'}=0$.  Recall
that it is such averaged quantities that we are interested in; the
admittance $\mu(\omega)$ is obtained by suitably averaging over
the ensemble, cf.\ \eqref{presponse0}.

Requiring the boundary condition \eqref{bc_forced_x=xc}, with electric field
\eqref{monochr_wvelflx} at $\rho=\rho_c$ and the condition that $\ev{A'}=0$,
we determine the average values of $A',B'$ to be
\begin{align}
 \ev{A'}=0,\qquad
 \ev{B'} =
 {2i\pi\ap\,\ell^4\over r_H^3}
 {1-i\nu\over \nu\,(1-i \rho_c \nu)}
 \left({\rho_c-1\over \rho_c+1}\right)^{i\nu/2}\,E_0  \ .
 \label{A'B'ave}
\end{align}
From this we infer that the average value of $X$ at the UV cut-off $\rho=\rho_c$ is
\begin{align}
 \ev{x(t)}
 =\ev{X(t,\rho_c)}=
 {2i\pi\ap\,\ell^4\over r_H^3}\,
 {1-{i\nu/ \rho_c}\over \nu \, (1-i\rho_c \nu)} \;
 E_0\, e^{-i\omega t} \ ,
\end{align}
which in turn implies that the average value of the momentum $p=m\dot x$
is
\begin{align}
 \ev{p(t)}
 =
 {2\pi\ap\ell^4 \,m\omega \over r_H^3}\,
 {1-{i\nu/ \rho_c}\over \nu \,(1-i\rho_c \nu)}\;
 E_0 \,e^{-i\omega t} =
 {\ap\beta^2\, m \over 2\pi\,\ell^2}\,
 {1-{i\nu/ \rho_c}\over 1-i\rho_c \nu}
 \;E_0\, e^{-i\omega t} \ .
\end{align}
Comparing this with
\eqref{presponse0}, we obtain the admittance
\begin{align}
 \mu(\omega)=
 {1\over \gamma[\omega]-i\omega}
 =
 {\ap\beta^2\, m \over 2\pi\,\ell^2}\,
 {1-{i\nu/ \rho_c}\over 1-i\rho_c \nu}.
 \label{boundary_mu}
\end{align}

%
%
%
%

A simple check on the consistency of these computations
is to compute the energy flow along the string falling into the
horizon, which must be equal to the work done by the external force.
For the theory \eqref{SNG}, the stress-energy tensor is
\begin{align}
 T^{\mu}_\nu=
 {1\over 2\pi \,\ap}\left(g^{\mu\kappa}\,\delta^\lambda_\nu
 -{1\over 2}\,\delta^{\mu}_\nu \,g^{\kappa\lambda}\right)
 G_{IJ}\,\partial_\kappa X^I \partial_\lambda X^J.
\end{align}
Because we are working in static gauge this world-sheet stress-energy
tensor measures the spacetime energy.
In the case of the BTZ spacetime \eqref{nonrotBTZ}, the flow of energy
along the $r$ direction is
\begin{align}
 \sqrt{-g}\,T^r_t
 ={r^2\,(r^2-r_H^2) \over 2\pi \,\ap \,\ell^4} \Re[\partial_r \overline{X}\, \partial_t X]
 ={r_H^3\, \rho^2(\rho^2-1) \over 2\pi \,\ap \,\ell^4}\, \Re[\partial_\rho \overline{X}\partial_t X].
\label{energyflow}
\end{align}
Here, we replaced $\partial_\kappa X\partial_\lambda X\to \Re[\partial_\kappa
\overline{X}\partial_\lambda X]$ so as to work directly with complex fields.
Consider the solution for  $X(t,\rho)$ as in \eqref{X(t,r)_forced}
with the coefficients $A',B'$ given by the average value
\eqref{A'B'ave} (we ignore thermal fluctuations in replacing the
amplitudes by their average).  Then, \eqref{energyflow} evaluates
to\footnote{The plus sign is because $t$ is a lower index.  If we
raise $t$, this will have a minus sign, indicating a flow of
energy toward the direction of the horizon (smaller $r$).}
\begin{align}
 \sqrt{-g}\,T^r_t
 ={2\pi \,\ap\,\ell^2 \,E_0^2\over r_H^2}
 {1+\nu^2\over 1+\rho_c^2\,\nu^2} \ .
 \label{energyflow_expl}
\end{align}
On the other hand, the work done per unit time (namely, power) by the
electric field $E$ acting on the endpoint at $X(t,\rho_c)$ is
\begin{align}
 \Re[{\overline{E}}\, \partial_t X(t,\rho_c)] \ ,
\end{align}
where $E$ is given by \eqref{monochr_wvelflx}.  For $X(t,\rho)$ in
\eqref{X(t,r)_forced}, it is easy to check that this equals
\eqref{energyflow_expl}.
Hence indeed as expected, the work done by the external force is
transmitted down the string into the black hole horizon and the energy is
thus dissipated away.

\subsection{The holographic auto-correlation function and time scales}
\label{sec:HolAuto}

We now turn to the computation of the random force correlator,
$\kappa(\omega)$.  From the $\ev{x\,x}$ correlator \eqref{xx_reg}, we can
compute the $\ev{p\,p}$ correlator as
\begin{align}
  \ev{{:\!p(t)\,p(0)\!:}}
 = -m^2\,\partial_t^2 \ev{{:\!x(t)\,x(0)\!:}}
 &=
 {\ap\,\beta^2 \,m^2 \over \pi^2\, \ell^2}
 \int_0^\infty \!\! {d\omega}\,
 {1+\nu^2\over 1+\rho_c^2 \nu^2}\,
  {\omega \cos{\omega t}\over e^{\beta \omega}-1}\notag\\
 &
 =
 {\ap\,\beta \,m^2 \over \pi \,\ell^2}
 \int_{-\infty}^\infty {d\omega\over 2\pi}\,
 {1+\nu^2\over 1+\rho_c^2\nu^2}\,
  {\beta|\omega|\,e^{-i\omega t}\over e^{\beta |\omega|}-1}.
 \label{pp_reg}
\end{align}
To obtain the power spectrum defined in \eqref{pwrspctr_def} for the
momentum $p$ we Fourier transform in time $t$ to obtain
\begin{align}
 I_p^{\text{n}}(\omega)
 & =
 {\ap\beta m^2 \over \pi \ell^2}\,
 {1+\nu^2\over 1+\rho_c^2\nu^2}\,
  {\beta|\omega|\over e^{\beta |\omega|}-1}.
\end{align}
Here, the superscript ``n'' is for remembering that this power
spectrum was computed using the normal ordered correlator
$\ev{{:\!p\,p\!:}}$.  Then we can exploit the relation \eqref{Ip_and_IR} between the power spectrum for the auto-correlation function and the momentum spectrum and the  previously derived  expression for $\mu(\omega)$, \eqref{boundary_mu}, to obtain the power
spectrum for the random force $R$, which is nothing but the random force
correlator $\kappa^{\text{n}}(\omega)$:
\begin{align}
\kappa^{\text{n}}(\omega)
 =I_R^{\text{n}}(\omega)
 & =
 {I_p^{\text{n}}(\omega)\over |\mu(\omega)|^2}
 =
 {4\pi \ell^2\over \ap\beta^3}\,
 {1+\nu^2 \over 1+\rho_c^2 \nu^2}\,
 {\beta|\omega|\over e^{\beta|\omega|}-1}.
\label{kappa^n(omega)}
\end{align}

Next, let us compute the physical time scales $t_{\text{relax}}$ and
$t_{\text{coll}}$.  First, from \eqref{boundary_mu}, one can compute the
relaxation time $t_{\text{relax}}$ defined in \eqref{t_relax_gen} as:
\begin{align}
 t_{\text{relax}}&=\mu(\omega=0)\sim {\ap\beta^2 m\over \ell^2}.
\end{align}
To compute the collision duration time $t_{\text{coll}}$, we first need
the real time auto-correlation function for the random force $\ev{R\,R}$:
\begin{align}
 \kappa^{\text{n}}(t)
 =\ev{{:\!R(t)R(0)\!:}}
 =\int_{-\infty}^\infty \,{d\omega\over 2\pi} \,I_{R}^{\text{n}}(\omega)\,e^{-i\omega t}.
\end{align}
By using the explicit form \eqref{kappa^n(omega)}, we obtain
\begin{align}
 \kappa^{\text{n}}(t)
 ={2\,\ell^2 \over \ap \,\beta^4}
 \left[ \rho_c^2 \,h_1(t,\beta)-(\rho_c^2-1)\, h_2(t,\beta,\rho_c) \right],
\end{align}
where we defined the functions
\begin{align}
 h_1(t,\beta)\equiv \int_{-\infty}^\infty \!\! dx\, {|x|\, e^{-itx/\beta}\over e^{|x|}-1},
 \qquad
 h_2(t,\beta,\rho_c)\equiv \int_{-\infty}^\infty \!\! dx\,
 {|x|\, e^{-itx/\beta}\over (1+({x\over 2\pi \rho_c})^2)(e^{|x|}-1)}.
\end{align}
For $\rho_c\gg 1$, $h_1$ and $h_2$ are almost equal; if $x\ll \rho_c$,
we can approximate $1+({x\over 2\pi \rho_c})^2$ in the integrand of
$h_2$ by $1$ while, if $x\gtrsim \rho_c\gg 1$, the integrand is almost
vanishing because of the Bose--Einstein like factor $1/(e^{|x|}-1)$.
Therefore, the $RR$ correlator evaluates to
\begin{align}
 \kappa^{\text{n}}(t)
 \approx {2\ell^2\over \ap \beta^4}h_1(t,\beta)
 = {2\ell^2\over \ap \beta^4}
 \biggl[\biggl({\beta\over t}\biggr)^2-{\pi^2\over \sinh^2(\pi t/\beta)}\biggr].\label{kappa^n}
\end{align}
This function has a support of width of order $\beta$ around $t=0$.  Therefore,
using  \eqref{def_t_coll}  we obtain the collision duration time
\begin{align}
 t_{\text{coll}}\sim {\beta}={1\over T},
\end{align}
The $T$ dependence is as it should be from dimensional analysis in a CFT
at temperature $T$, but the fact that this is independent of the 't
Hooft coupling $\lambda$ is not trivial.

The ratio of the two time scales is given by
\begin{align}
 {t_{\text{relax}}\over t_{\text{coll}}}
 \sim {\ap m\over \ell^2 T}\sim {m\over\sqrt{\lambda}\,T},
\end{align}
where we related $\alpha'/\ell^2$ to the boundary 't Hooft
coupling\footnote{Strictly speaking, this is only the 't Hooft coupling
$\lambda$ in the standard AdS${}_5$ case, but we will use the same
terminology to denote $\ell^4/\ap^2$ for other values of $d$ as well.}
by using the relation $\ell^4/\ap^2\sim \lambda$
\cite{Herzog:2006gh, CasalderreySolana:2006rq}. In the weak or
moderate coupling regime, $\lambda\lesssim 1$, we can make this ratio
large by considering a Brownian particle with $m\gg T$ and obtain the
standard Brownian motion as explained below \eqref{tr>>tc}; the Brownian
particle becomes thermalized only after numerous collisions with fluid
particles.  In the strong coupling regime, $\lambda\gg 1$, however, this
is not the case and, in order to have the standard picture, we have to
consider a much heavier Brownian particle with mass $m\gg
\sqrt{\lambda}\,T$, which is always possible.  On the other hand, if
$T\ll m\ll \sqrt{\lambda}\,T$, the situation is totally
different. Although the effect of a collision with a single fluid
particle (with energy $\sim T$) is small, because the Brownian particle
interacts with many fluid particles at the same time, it can become
thermalized in a time much shorter than the time it takes for a single
process of collision.
To make this claim more quantitative, we estimate in
Appendix~\ref{sec:t_mfp} the average time $t_{\text{mfp}}$ between
collisions. The contribution that a single collision makes to the
random force $R(t)$ has width $t_{\rm coll}$.  $R(t)$ consists of
many such contributions, with the typical distance in time between
two collisions being $t_{\rm mfp}$. Determining $t_{\rm mfp}$ is
not entirely straightforward, as it requires us to analyze the
four-point correlation function of the random force, and we only
find a non-trivial answer once we take the fourth order correction
to the Nambu--Goto action into account. As a result,
$t_{\text{mfp}}$ is suppressed by a factor of $1/\sqrt{\lambda}$
compared to $t_{\text{coll}}$, the final result being
\begin{align}
 t_{\rm mfp}\sim {1\over \sqrt{\lambda}\,T}.
\end{align}
At weak coupling $\lambda\ll 1$, we have $t_{\rm mfp}\gg t_{\rm coll}$
and the standard kinetic theory picture, where the Brownian particle is
occasionally hit by a fluid particle, is valid.  On the other hand, at
strong coupling $\lambda\gg 1$, we have $t_{\rm mfp}\ll t_{\rm coll}$,
namely, many collisions occur within the time scale for a single
collision process to take place.\footnote{Perhaps the term ``mean free
path time'' is not an appropriate one in this regime where a second
collision takes place before the first one ends, and thus the particle
is never freely moving.  However, there being no other choice, we will
continue to use this term in the strongly coupled regime.} This supports
the picture above that the Brownian particle interacts with many fluid
particles at the same time.
%

\section{Fluctuation-dissipation theorem}
\label{sec:FDthm}

Thus far we have seen how the string probe in the bulk geometry
holographically captures the Brownian motion of an external test
particle introduced in the boundary CFT plasma. As we have seen
explicitly, one can derive the Langevin equation for the string endpoint
by tracing back the information about the part of the string that is
touching the black hole and hence gets thermally excited due to the
outgoing Hawking quanta. One of the hallmarks of non-equilibrium
statistical mechanics is the fluctuation-dissipation theorem
\cite{Kubo:f-d_thm, KTH} which relates the observables in the system
perturbed infinitesimally away from equilibrium to equilibrium
quantities.  We now turn to show that not only are the results we
derived in section \ref{sec:semiclass} consistent with the
fluctuation-dissipation theorem, but that we can in fact obtain this
result directly from the gravity side.

\subsection{Linear response theory}
\label{sec:LinResp}
We begin our discussion of the fluctuation-dissipation theorem
with a lightning review of linear response theory
\cite{Kubo:f-d_thm, KTH}.

Consider a system whose unperturbed Hamiltonian is given by
$H$. Assume that, in the infinite past $t=-\infty$, the system was in an
equilibrium state with the density matrix
\begin{align}
 \rho_e={e^{-\beta H}\over \tr e^{-\beta H}}.
\end{align}
Now perturb the system by adding an external force $K(t)$ conjugate to
a quantity $A$.  The total Hamiltonian is
\begin{align}
 H_{\text{tot}}&=H+H_{\text{ext}}(t)=H-A\, K(t).\label{H_tot}
\end{align}
Under this perturbation, the change in another quantity $B$ is
given, to the first order in the perturbation $H_{\text{ext}}$, by
the so-called Kubo formula:
\begin{align}
 \Delta B(t)&=
 \int_{-\infty}^t dt'\, K(t')\,\phi_{BA}(t-t'),\qquad
 \phi_{BA}(t)\equiv -i\ev{[A(0),B(t)]},
 \label{Kubo_formula}
\end{align}
where we defined $\ev{\CO}\equiv\tr(\rho_e \CO)$ and $\CO(t)=e^{iHt}\CO
e^{-iHt}$.  The function $\phi_{BA}(t)$ is called the response
function\index{response function}.

If we consider a periodic force with frequency
$\omega$,
\begin{align}
 K(t)=K_0 \,e^{-i\omega t},
\end{align}
then \eqref{Kubo_formula} gives the following change in $B$:
\begin{align}
 \Delta B(t)&=\mu_{BA}(\omega) \, K_0\,  e^{-i\omega t},
\end{align}
where the admittance $\mu_{BA}(\omega)$ is given by
\begin{align}
 \mu_{BA}(\omega)=
 \int_{0}^\infty dt\,\phi_{BA}(t)\,e^{i\omega t}
 ={1\over i}\int_{0}^\infty dt\,\ev{[A(0),B(t)]}e^{i\omega t}
 =\beta \int_0^\infty dt\,\ev{\dot A(0); B(t)}e^{i\omega t},
\label{FDthm}
\end{align}
and with the \emph{canonical correlator} $\ev{X;Y}$ defined by
\begin{align}
 \ev{X;Y}
 ={1\over\beta}\int_0^\beta d\lambda\,\ev{e^{\lambda H}Xe^{-\lambda H}Y}
 ={1\over\beta}\int_0^\beta d\lambda\,\ev{X(-i\lambda)\,Y}  \ ,
\label{def_can_corr}
\end{align}
which satisfies the following properties
\begin{align}
 \ev{X(0);Y(t)}=\ev{Y(t);X(0)}=\ev{Y(0);X(-t)}.
\label{cancorr_prop}
\end{align}
The relation \eqref{FDthm} is called the  fluctuation-dissipation
theorem, because the right hand side is the fluctuation
(correlator) in the equilibrium state $\rho_e$, while the left
hand side yields the admittance which is related to the
dissipation (friction).

In the case of Brownian motion, we can take $A=x$ and
$H_{\text{ext}}=-xK(t)$, where $K(t)$ is identified with the external
force appearing in the Langevin equation \eqref{genLE}.  Then, for
$B=p$, we obtain the admittance
\begin{align}
 \mu(\omega)
 ={\beta\over m}\int_0^\infty dt\, \ev{p(0);p(t)}\, e^{i\omega t}.
\label{FDthm_BM}
\end{align}
Due to the relations \eqref{cancorr_prop}, this implies
\begin{align}
 2\Re\mu(\omega)
 ={\beta\over m}\int_{-\infty}^\infty dt\, \ev{p(0);p(t)}\, e^{i\omega t}
 ={\beta\over m}\, I_{p}^{\text{c}}(\omega),
\label{FDthm_BM2}
\end{align}
where $I_{p}^{\text{c}}(\omega)$ is the power spectrum for $p$ defined
using the canonical correlator.
From this, using the relation \eqref{Ip_and_IR}, one can derive a more
direct relation between the friction and random force as
\begin{align}
 2\Re\gamma(\omega)
 ={\beta\over m}\, I_R^{\text{c}}(\omega)
 ={\beta\over m}\, \kappa^{\text{c}}(\omega),
 \label{FDthm2_BM}
\end{align}
which is sometimes called the second fluctuation-dissipation theorem, in
contrast with \eqref{FDthm_BM} or \eqref{FDthm_BM2} which is sometimes
called the first fluctuation-dissipation theorem \cite{Kubo:f-d_thm,
KTH}.

\subsection{Explicit check of fluctuation-dissipation theorem}
\label{sec:fdcheck}

The fluctuation-dissipation relations \eqref{FDthm_BM},
\eqref{FDthm_BM2}, and \eqref{FDthm2_BM} for Brownian motion were
derived from the field theory viewpoint and are not immediately obvious
from the bulk viewpoint.  Here, let us explicitly check that they indeed
hold using the explicit results obtained from the bulk in section
\ref{sec:semiclass}.

Similarly to \eqref{xx} or \eqref{xx_reg}, we can compute the
canonical correlator for $x$ as
\begin{align}
  \ev{x(0);x(t)}
 &=
 {\ap\beta\over \pi \ell^2}\int_{-\infty}^\infty {d\omega\over2\pi}\,
 {1\over\omega^2}\,
 {1+\nu^2 \over 1+\rho_c^2 \nu^2}e^{-i\omega t}.
\label{xx_can}
\end{align}
Because $p=m\dot x$, this implies the following canonical correlator for
$p$:
\begin{align}
  \ev{p(0);p(t)}
 &=
 {\ap\beta m^2\over \pi \ell^2}\int_{-\infty}^\infty {d\omega\over2\pi}\,
 {1+\nu^2 \over 1+\rho_c^2 \nu^2}e^{-i\omega t}.
\label{pp_can}
\end{align}
This means that the power spectrum for $p$ is
\begin{align}
 I_p^{\text{c}}(\omega)
 =
 {\ap\beta m^2\over \pi \ell^2}
 {1+\nu^2 \over 1+\rho_c^2\nu^2}.
\label{FDthm_BM3_RHS}
\end{align}
On the other hand, from \eqref{boundary_mu}, we immediately obtain
\begin{align}
 2\Re\mu(\omega)&=
 {\ap \beta^2 m\over \pi \ell^2}\,
 {1+\nu^2 \over 1+\rho_c^2\nu^2}.
\label{FDthm_BM3_LHS}
\end{align}
By comparing \eqref{FDthm_BM3_RHS} and \eqref{FDthm_BM3_LHS}, we see
that the fluctuation-dissipation theorem of the form \eqref{FDthm_BM2}
indeed holds.
This also implies the second fluctuation-dissipation theorem
\eqref{FDthm2_BM}.

These relations can be regarded as providing evidence that the motion of
the string endpoint in the bulk can be described by a generalized
Langevin equation.  Note, in particular, that the way we derived the
fluctuation (correlator) and the way we derived dissipation (admittance)
were very different; for the former, we assumed thermal Hawking
radiation near the horizon and measured the position of the string
endpoint at the UV cut-off while, for the latter, we considered forced
motion imposing a boundary condition at the horizon which was
essentially the purely ingoing boundary condition.
In the next subsection we describe how these two quantities are related
directly from the bulk point of view. However, it would be desirable to
have a more intuitive physical understanding of why this should be the
case.

\subsection{Bulk proof of fluctuation-dissipation theorem}
\label{sec:bullkfd}

In subsection \ref{sec:fdcheck}, we demonstrated that the
fluctuation-dissipation relations holds for the special case of string
probes in the BTZ spacetime by an explicit bulk computation. We now
prove that the fluctuation-dissipation relations hold more generally,
again from the bulk viewpoint.

Consider a string probe in the $d$-dimensional metric \eqref{genmetric}.
We would like to turn on an electric field $F_{It}=E_I(t)$ on the flavor
D-brane at $r=r_c$ and consider the resulting position
$x^I(t)=X^I(t,r_c)$ of the string endpoint in response to it.
If we take $A_t=E_I(t) \, X^I$, $A_I=0$, then the boundary action,
\eqref{Sbndy}, can be written as
\begin{align}
 S_{\text{bdy}}&
 =\int dt\, E_I(t)X^I
 =\int dt\,dr\, \delta(r-r_c) E_I(t)X^I.\label{S_bndy_src}
\end{align}
This can be regarded a source term for the field $X^I$; upon inclusion
of this term, the equation of motion \eqref{eom_NG} is changed to
\begin{align}
 \nabla^\mu\left[G_{IJ}(x)\, \partial_\mu X^I(x)\right]
 =-{2\pi\ap\over\sqrt{-g}}\, \delta(r-r_c) \, E_I(t).\label{eom_NG_src}
\end{align}
As is standard, we can solve this by using the retarded propagator
\begin{align}
 D_{\text{ret}}^{IJ}(t,r|t',r')
 =\theta(t-t')\ev{[X^I(t,r),X^J(t',r')]},
\label{D_ret}
\end{align}
where $X^I(t,r)$ satisfies the equation of motion \eqref{eom_NG} (or equivalently
\eqref{eom_NG_src} with the right hand set to zero) and can be expanded
in modes as in \eqref{X_expn}. Namely,
\begin{align}
\begin{split}
  X^I(t,r)&=\sum_{\omega>0} [u_\omega^I(t,r)a_\omega+u_\omega^I(t,r)^*a_\omega^\dagger],\\
 [a_\omega,a_{\omega'}]&=[a_\omega^\dagger,a_{\omega'}^\dagger]=0,\qquad
 [a_\omega,a_{\omega'}^\dagger ]=\delta_{\omega\omega'},
\end{split}
\label{X^I_expn_gen}
\end{align}
where $\{u_\omega^I(t,r)\}$ is a normalized basis of solutions to
\eqref{eom_NG}.
Note that, with
\eqref{X^I_expn_gen}, the commutator appearing in $\eqref{D_ret}$ is
actually a c-number and $D_{\text{ret}}^{IJ}$ is independent of the
state with respect to which we take the expectation value.
$D_{\text{ret}}^{IJ}$ can be shown to satisfy
\begin{align}
 \nabla^\mu\left[G_{IJ}(x)\, \partial_\mu D_{\text{ret}}^{JK}(t,r|t',r')\right]
 =i \, {2\pi\ap\over\sqrt{-g}} \, \delta^K_I\, \delta(t-t')\, \delta(r-r').
\end{align}
Therefore, the solution to \eqref{eom_NG_src} can be written, using
$D_{\text{ret}}^{IJ}$, as
\begin{align}
\begin{split}
  X^I(t,r)
 &=i\int_{-\infty}^\infty  dt'\, D_{\text{ret}}^{IJ}(t,r|t',r_c)E_J(t')\\
 &=i\int_0^\infty dt''\,\ev{[X^I(t,r),X^J(t-t'',r_c)]}E_J(t-t'').
\end{split}
\end{align}
By setting $r=r_c$, we obtain the position $x^I(t)=X^I(t,r_c)$ of the
string endpoint in response to the external force $E_I(t)$ as:
\begin{align}
  x^I(t)
 &=i\int_0^\infty dt''\,\ev{[x^I(t),x^J(t-t'')]}E_J(t-t'').
 \label{x_respns_bulk}
\end{align}

If the fluctuation-dissipation theorem is to hold, this must be equal to
the $x^I(t)$ obtained by using the Kubo formula \eqref{Kubo_formula},
identifying $x^I(t)$ with the position of the Brownian particle in the boundary.
From the action \eqref{S_bndy_src}, one reads off the external force
appearing in \eqref{H_tot} to be $AK=E_J x^J$.  Then, by applying the Kubo
formula \eqref{Kubo_formula} for $B=x^I$,
\begin{align}
\begin{split}
  x^I(t)&=-i\int_{-\infty}^t dt'\, E_J(t') \ev{[x^J(0),x^I(t-t')]}\\
 &=i\int_{0}^{\infty} dt''\, \ev{[x^I(t''),x^J(0)]} E_J(t-t''),
\end{split}
 \label{x_respns_bndy}
\end{align}
where $t''=t-t'$.
Because the system is stationary, the expectation value is invariant
under shift of time: $\ev{[x^I(t''),x^J(0)]}=\ev{[x^I(t),x^J(t-t'')]}$.
Therefore, the bulk response \eqref{x_respns_bulk} is the same as the
expression \eqref{x_respns_bndy} computed from the boundary Kubo
formula.
This implies that the fluctuation-dissipation relations
\eqref{FDthm_BM}, \eqref{FDthm_BM2}, \eqref{FDthm2_BM} indeed hold.

\section{General dimensions}
\label{sec:gen_dim}

\def\be{\begin{equation}}
\def\ee{\end{equation}}

Thus far we have considered the case of $d=3$ dimensional
asymptotically AdS spacetimes only, which had the advantage that
the wave equation for the modes of the string was exactly
solvable. For general $d$, this is no longer possible and we have
to use approximate methods. In this section, we employ the low
frequency approximation $\omega\ll T$ and briefly summarize how
some of the results of the previous sections get modified in the
case of asymptotically AdS$_d$ spacetimes with general $d$.

The starting point is the metric (\ref{AdSdBH}), with Hawking
temperature given in (\ref{Hawk_temp_d}). The tortoise coordinate
$r_{\ast}$ is defined via
\be \label{deftor}
dr_{\ast}= \frac{\ell^2}{r^2\, h(r)}\, dr.
\ee
If we define $\eta= \exp[2\pi i/(d-1)]$, an explicit
expression for the tortoise coordinate is
\be
r_{\ast} = \sum_{k=0}^{d-2} \frac{\ell^2}{(d-1)\,\eta^k \,r_H}
\;\log\left({r\over r_H}-\eta^k\right).
\ee
The term with $k=0$ shows that near the horizon, $r_{\ast} \sim
\frac{\ell^2}{(d-1)\,r_H} \,\log\bigl({r\over r_H}-1\bigr)$, and from
(\ref{deftor}) we also see that the behavior near infinity is
\be
r_{\ast} \sim -\frac{\ell^2}{r}, \qquad r\rightarrow \infty .
\ee
The generalization of (\ref{phi_eom}) to arbitrary $d$ reads
\be \label{fundeq}
 -\partial_t^2 X + \frac{h(r)}{\ell^4} \,\partial_r [r^4\,h(r) \,\partial_r X] =0.
\ee
As before, we will exploit the translational invariance along $t$ to
decompose modes in plane waves; for convenience consider solutions of
the form
\be
X(t,r) = e^{-i\omega t}\, r^{-1} \,\Phi_{\omega}(r) \ ,
\ee
from which it follows that the functions $\Phi_\omega(r)$ satisfy the
following equation
\be
\left[ \frac{\partial^2}{\partial r_{\ast}^2} + \omega^2 - V(r)
\right] \Phi_{\omega}(r) =0\ ,
\label{fundeq2}
\ee
with
\be
V(r) = \frac{1}{\ell^4} \, r^2 \,h(r) \, \left[2 \,h(r) + r\, h'(r)\right].
\ee
The wave equation \eqref{fundeq2} can be thought of as a
time-independent Schr\"odinger equation for a particle moving in
potential $V(r)$.

As in section \ref{sec:semiclass}, we want to exploit the semiclassical
physics of Hawking radiation to learn about the behaviour of the string
endpoint on the boundary. Once again it is worth noting that the
dynamics of the scalar field $X(t,r)$ is similar to a minimally coupled
scalar field propagating in an asymptotically AdS$_2$ spacetime with an
event horizon. We would like to compute the admittance for the Langevin
equation in this case.

In order to redo the computation that led to (\ref{boundary_mu}) we need
to find the solution of the wave equation \eqref{fundeq2} which is
purely ingoing at the horizon $r=r_H$.  Let us denote this particular solution
of the wave equation by $X^{-}_{\omega}(r)$.  It is not possible to
obtain this for general frequencies $\omega$ and hence we employ a low
frequency approximation $\omega\ll T$ and use the so-called matching
technique.  Here, we only write down the final result of the
computations, relegating the details to Appendix \ref{app:gen_dim}\@.
The solution that is purely ingoing at the horizon behaves near infinity
as
\be
X^-_\omega(\rho)  = C^+\, X^+_C(\rho) + C^- \,X^-_C(\rho),
\label{X-omega}
\ee
where
\be
C^{\pm} = \frac{1}{2}\,\left(1\pm\frac{1}{\nu^2} +i\,b\,\nu\right),\qquad
X^{\pm}_C(\rho) = \left(1\mp \frac{i\nu}{\rho}\right) e^{\pm i\,\nu/\rho}.
\label{indepsol_C}
\ee
Here, $b$ is a constant independent of $\nu$, whose precise value
is not relevant for our purpose.  Also, as before, we defined dimensionless quantities
\be
\rho \equiv \frac{r}{r_H},\qquad \nu\equiv \frac{\ell^2\,  \omega}{r_H}.
\ee
In terms of these, the low frequency condition $\omega\ll T$ reads
$\nu\ll 1$.  Actually, this result \eqref{X-omega} is valid only
to leading order in the $\nu$ expansion.

By carefully redoing the calculation in subsection \ref{sec:Forcedm}, one
can show that there is the following relation between the ingoing mode
and the admittance $\mu(\omega)$: \be \mu(\omega) =
\frac{1}{\gamma[\omega]-i\omega}
= -\frac{i\,(d-1)^2 \,\alpha' \,m \,\beta^2 \,\nu}{8\pi\, \ell^2\,\rho_c^4} \,\frac{X^-_\omega(\rho_c)}{\p_{\rho_c}X^-_\omega(\rho_c)} \ .
\ee
Using the explicit expression of $X_\omega^-$ (Eqs.\ \eqref{indepsol_C},
\eqref{X-omega}), the final result is
%
\be \label{mu_d}
\mu(\omega)
= \frac{i(d-1)^2 \,\alpha'\, m \,\beta^2}{8\pi \,\ell^2\,\rho_c^2\, \nu}\,
\frac{\left[( 1+ i\, b\,\nu)\, \nu\,\rho_c - i \right] \, \nu \, \cos\bigl(\frac{\nu}{\rho_c}\bigr) +
i\, \left[\rho_c -i\,\nu^3\, (1+i\, b\,\nu) \right] \, \sin\bigl(\frac{\nu}{\rho_c}\bigr)}{(1+i\,b\,\nu)\, \nu^2 \, \cos\bigl(\frac{\nu}{\rho_c}\bigr) + i \, \sin\bigl(\frac{\nu}{\rho_c}\bigr)}.
\ee
As mentioned above, this result is valid only to leading order in
the expansion in $\nu$.
For small $\nu$, the right hand side of \eqref{mu_d} behaves as
\be
\mu(\omega)
=\frac{(d-1)^2\,\alpha'\,\beta^2 \,m }{8\pi\,\ell^2} +\CO(\omega).
\ee
For $d=3$, this agrees with (\ref{boundary_mu}) for $\nu\rightarrow
0$. Furthermore, this agrees with the drag force result
\eqref{gamma_AdSd} for general $d$, because $\gamma_0=\mu[0]^{-1}$.

Just as we did for the $d=3$ case in subsection \ref{sec:Brendpt},
we could also compute $\kappa(\omega)$ for general $d$ in the low
frequency approximation.  However, this is not necessary, because
we can directly obtain $\kappa(\omega)$ from the
fluctuation-dissipation theorem (\ref{FDthm2_BM}), whose validity
we already demonstrated for all values of $d$ in subsection
\ref{sec:bullkfd}.


%
%
%

\section{Stretched horizon and Brownian motion}
\label{sec:BM_on_strtch_hor}

\begin{figure}[htbp]
  \begin{quote}
 \begin{center}
  \epsfxsize=7.5cm \epsfbox{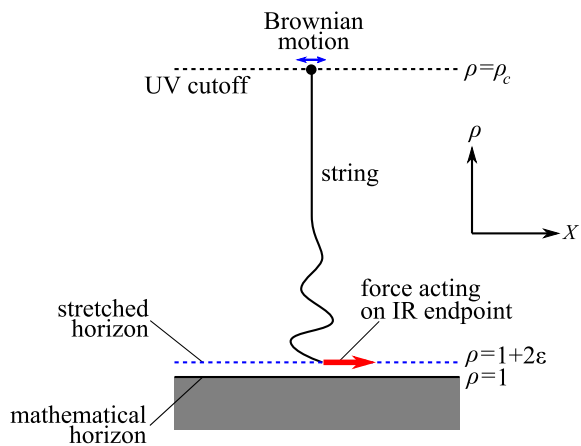} \caption{\sl A
 membrane-paradigm like picture of the Brownian motion.  There are
 friction and random force acting on the IR endpoint of the string on
 the stretched horizon, effectively giving the boundary condition.}
 \label{fig:membrane_paradigm}
\end{center}\end{quote}
\end{figure}

The main philosophy of the membrane paradigm \cite{Thorne:1986iy} is
that, as far as an observer staying outside a black hole horizon is
concerned, physics can be effectively described by assuming that the
objects outside the horizon are interacting with an imaginary membrane,
which is endowed with physical properties, such as temperature and
resistance, and is sitting just outside the mathematical horizon.
In section \ref{sec:semiclass}, we assumed that the Brownian motion of
the UV endpoint of a string was caused by the boundary condition we
impose at the horizon---all ingoing modes are falling in without being
reflected, while the outgoing modes are always thermally populated.
A curious question then is whether this boundary condition can be
reproduced, in the spirit of the membrane paradigm, by postulating
some interaction of the string with a membrane at the stretched
horizon just outside the actual horizon.  The interaction
necessarily assumes a stochastic character, so it is natural to
expect it to be described by a sort of Langevin equation. For a
schematic explanation, see Figure \ref{fig:membrane_paradigm}. It
must be noted that the physics of the stretched horizon has been
discussed in the AdS/CFT context previously in
\cite{Kovtun:2003wp,Saremi:2007dn,Fujita:2007fg,Starinets:2008fb}
and more recently in \cite{Iqbal:2008by} where there is a nice
discussion regarding the dynamics of the stretched horizon and the
universality of hydrodynamic coefficients. We will now turn to a
derivation of the properties of the stretched horizon in section
\ref{sec:LangStretch} and then proceed to ask whether we can learn
anything about the microscopic structure of the stretched horizon
in section \ref{sec:Granular}.

\subsection{Langevin equation on stretched horizon}
\label{sec:LangStretch}

Let us consider placing an imaginary ``IR brane'' near the horizon
at $\rho_s=1+2\,\epsilon$, $\epsilon\ll 1$ and assume that the
string ends on it.\footnote{In reality a string that dips into the
black hole will continue merrily past the horizon without any
trouble; the quickest way to see this of course is to pass to
coordinates that are regular on the horizon such as ingoing
Eddington-Finkelstein or Kruskal coordinates. Here we are
interested in mimicking the boundary conditions of the black hole
and hence will postulate there to be an imaginary boundary in the
IR at $\rho = \rho_s$.}  If we assume that a force is acting on
the endpoint, the equation of motion for the endpoint is, just as
in \eqref{bc_forced_x=xc}, given by
\begin{align}
 -{2 \,r_H^3\,\epsilon\over \pi\, \ap\,\ell^4}\;\partial_\rho X|_{\rho_s}=F_s^X,
\label{EOM_IRendpoint}
\end{align}
where $F_s^X$ is the force along the $X$ direction measured with respect
to the time $t$.  Note that there is no term like $m\,\ddot{X}$ on the
left hand side, because the endpoint has zero mass, having zero length.
We assume that the force $F_s^X$, just as in the usual Langevin equation
\eqref{genLE}, has frictional and stochastic components:
\begin{align}
 F_s^X(t)
 &=-\int_{-\infty}^{t} dt' \, \gammat(t-t') \, \partial_t X(t',\rho_s)
 +\Rt(t),\label{jxkt27Oct08}\\
 \ev{\Rt(t)}&=0,\qquad \ev{\Rt(t)\Rt(t')}=\kappat(t-t'),
\end{align}
where we allow the friction to depend on the past history through a
memory kernel $\gammat$.  We would like to choose $\gammat$ and
$\kappat$ appropriately to reproduce the correct boundary condition described above.

Near the horizon the fluctuation of the string is given by \eqref{X_near_hor}
\begin{align}
 X(t,\rho)
 & =\sum_{\omega>0}\sqrt{\ap\,\beta\,\over 2\,\ell^2\,\omega \,\log(1/\epsilon)}
 \left[ a_\omega^{(+)} \,e^{-i\omega(t-r_*)}+ a_\omega^{(-)}\, e^{-i\omega(t+r_*)}+{\rm h.c.}\right],
 \label{Xexpn_strtch}
\end{align}
where $\omega$ is discretized with $\Delta \omega$ given in
\eqref{Deltaomega}. $a_\omega^{(+)}$ and $a_\omega^{(-)}$ are
annihilation operators for outgoing and ingoing modes, respectively.
Depending on the boundary condition one imposes at the UV cut off,
$a_\omega^{(+)}$ and $a_\omega^{(-)}$ get related to each other (for
example, in the case of the Neumann boundary condition we imposed in
subsection \ref{ss:BdyCut}, they are related as $a_\omega^{(-)} =
e^{i\,\theta_\omega} \, a_\omega^{(+)}$). However, because we are
considering a Langevin equation which holds independent of such
relations, we regard $a_\omega^{(+)}$ and $a_\omega^{(-)}$ as
independent variables.

Plugging \eqref{Xexpn_strtch} in and going to the frequency space, we
 can write the equation of motion \eqref{EOM_IRendpoint} as
\begin{align}
 -i\,\sqrt{\ap\,\beta\,\omega \over 2\,\ell^2\,\log(1/\epsilon)}
 \left[
  \left(\gammat[\omega]+{r_H^2\over 2\pi\, \ap \,\ell^2}\right)\, a_\omega^{(+)}\, e^{i\omega r_*}
 +\left(\gammat[\omega]-{r_H^2\over 2\pi \,\ap \,\ell^2}\right)\,a_\omega^{(-)}\,e^{-i\omega r_*}
 \right]
 =\Rt(\omega)
 \label{EOM_endpt_Frr}
\end{align}
for $\omega>0$.  Here, $\gammat[\omega]$ is the Fourier--Laplace
transform of $\gammat(t)$ similar to \eqref{FL_trfm} while $\Rt(\omega)$
is the Fourier transform of $\Rt(t)$ as in \eqref{Fourier_trfm}.
In order to realize the boundary condition that all ingoing modes fall
in without reflection, we should set the coefficient of $a^{(-)}_\omega$
to zero (since we want to be able to set the ingoing amplitude
$a^{(-)}_\omega$ to any value).  This gives
\begin{align}
 \gammat[\omega]={r_H^2\over 2\pi \,\ap \,\ell^2}={2\pi\, \ell^2\over \ap \,\beta^2}
 \quad \Rightarrow \quad
 \gammat(t)={4\pi \,\ell^2\over \ap \,\beta^2}\,\delta(t).
\end{align}
Substituting this back into \eqref{EOM_endpt_Frr}, we obtain the
relation between the random force and the outgoing mode coefficients
$a_{\omega}^{(+)}$ as
\begin{align}
 \Rt(\omega)=-i\sqrt{8\pi^2\,\ell^2 \,\omega\over \ap\,\beta^3\,\log(1/\epsilon)}
 \, e^{i\omega r_*}\,a^{(+)}_\omega.\label{Rt_and_a^(+)}
\end{align}

If this random force is to realize the thermal nature of the
outgoing modes, $\ev{a_{\omega}^{(+)}{}^\dagger
a_{\omega}^{(+)}}=(e^{\beta \omega}-1)^{-1}$, then from
\eqref{Rt_and_a^(+)} we obtain
\begin{align}
 \CD(\omega)\,
 \ev{\Rt(\omega)^\dagger\, \Rt(\omega)}
& ={2\,\ell^2\,\omega \over \ap \,\beta^2 \,(e^{\beta\omega}-1)}  \label{RtRt_omega_corr0}
 \\
 &\approx {2\,\ell^2\over \ap\,\beta^3} \ ,
 \qquad {\rm for}\; \beta\, \omega\ll 1 \ .
 \label{RtRt_omega_corr}
\end{align}
Here $\CD(\omega)$ is the density of states defined in
\eqref{DoFomega}.  This means that the correlator for the random force
is
\begin{align}
 \kappat(t-t')=\ev{R(t)R(t')}
 =
 \int_{-\infty}^\infty d\omega\,\CD(\omega)\,
 \ev{\Rt(\omega)^\dagger\, \Rt(\omega)}\, e^{i\omega (t-t')}
 \approx {4\pi\,\ell^2\over \ap\,\beta^3}\,\delta(t-t').\label{kjvw27Oct08}
\end{align}
The delta function behavior is due to the approximation we made in
\eqref{RtRt_omega_corr}; the actual $\kappat(t-t')$ is nonvanishing for
$|t-t'|\lesssim \beta$, as one can see if one uses the original exact
expression \eqref{Rt_and_a^(+)}, \eqref{RtRt_omega_corr0}.

In summary, the boundary condition near the horizon can be effectively
realized by a Langevin equation for the string endpoint $X(t,\rho_s)$ at
the stretched horizon given by\footnote{Note that these relations are
operator relations whose full structure has been given in
\eqref{Rt_and_a^(+)} and \eqref{RtRt_omega_corr0}.}
\begin{align}
 -{2 \,r_H^3\,\epsilon\over \pi\,\ap\, \ell^4}\;\partial_\rho X
 =-{2\pi\,\ell^2\over \ap\,\beta^2}\;\partial_t X
 +\Rt(t),\qquad
\ev{\Rt(t)\Rt(t')}
 \approx {4\pi\,\ell^2\over\ap\,\beta^3}\;\delta(t-t').\label{LE_NH}
\end{align}
%
The two terms on the right hand side of the first equation are,
respectively, i) friction which precisely cancels the ingoing waves, and
ii) random force which is responsible for the outgoing modes being
thermally excited at the Hawking temperature.

Given the auto-correlation function for the random force $R_s(t)$
acting on the string endpoint at the stretched horizon, we can
exploit the Sutherland--Einstein relation \eqref{diff_const_def} to
compute the diffusion constant on the stretched horizon. We find
\begin{equation}
 D^s_{\text{AdS$_3$}} = \frac{2 \, T^2}{\kappa_s(0)} = \frac{\alpha'}{2 \, \pi \, \ell^2 \, T}
\label{Dstretched}
\end{equation}
which is the same as the diffusion constant for the string
endpoint undergoing Brownian motion in the boundary
\eqref{diff_const_AdS3}.  In deriving \eqref{Dstretched} we had to
assume that the dynamics of the string endpoint on the stretched
horizon obeys the Sutherland--Einstein relation derived for a point
particle.
In other words, we assumed that a point particle fixed on the
stretched horizon will experience the same friction and random
force as the ones appearing on the right hand side of
\eqref{LE_NH}, and thus will random walk with the diffusion
constant \eqref{Dstretched}.
In fact, we will now argue that this is not quite unexpected from
the viewpoint of the membrane paradigm.

In the context of the membrane paradigm, it is conventional to
ascribe transport properties to the stretched horizon. In fact, it
is well known that the shear viscosity of the black hole membrane
saturates the famous bound derived in the boundary field theory,
$\eta/s = 1/4\pi$, cf.,
\cite{Thorne:1986iy,Parikh:1997ma}.\footnote{See also
\cite{Eling:2006aw} for another membrane paradigm inspired
perspective on the ratio $\eta/s$.}  In the hydrodynamic regime of
the AdS/CFT correspondence, Ref.\ \cite{Iqbal:2008by} argued that
one can derive the universality of this ratio using the membrane
paradigm, {\it i.e.}, the physics of the stretched horizon similar
to the discussion given above. We have here focussed on the
stochastic Langevin process and derived the features of the
membrane that reproduce the physics of strings impinging on the
black hole. Again we see that the diffusion coefficient of heavy
quarks in the boundary \eqref{diff_const_AdS3} agrees with that
derived for the stretched horizon \eqref{Dstretched}.

\subsection{Granular structure on the stretched horizon}
\label{sec:Granular}

In \cite{Susskind:1993ws, Halyo:1996vi}, Susskind and collaborators put
forward a provocative conjecture that a black hole is made of a
fundamental string covering the entire horizon.  Although this picture
must be somewhat modified \cite{Horowitz:1996nw} since we now know that
branes are essential ingredients of string theory, it is still an
attractive idea that, in the near horizon region where the local
temperature becomes string scale, a stringy ``soup'' or ``cloud'' of
strings and branes is floating around, covering the entire horizon.

\begin{figure}[tbp]
  \begin{quote}
 \begin{center}
  \epsfxsize=4cm \epsfbox{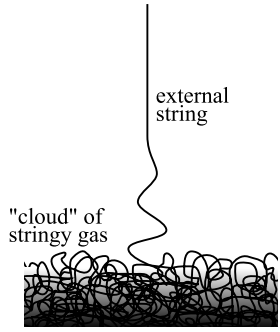} \caption{\sl A possible
 microscopic picture of a black hole, where the horizon is covered by a
 stringy ``cloud'' made of strings and branes. An external fundamental
 string ending on a horizon is dissolved into the cloud and incessantly
 kicked around by the cloud.}  \label{fig:clouds}
\end{center}\end{quote}
\end{figure}

If this picture is true, one naturally expects that it is this
stringy cloud that is exerting frictional and stochastic forces on
the IR endpoint of the fundamental string as described by
\eqref{LE_NH}; see Figure \ref{fig:clouds}.  Can we
learn anything about this stringy cloud?
The stretched horizon is located a distance $\sim l_s=\sqrt{\ap}$ away
from the mathematical horizon \cite{Halyo:1996vi}. It is occupied by a
string of length $L\sim S\, l_s$, with $S$ the entropy of the black
hole. If we associate one degree of freedom to each string segment of
length $l_s$, the number of degrees of freedom equals the entropy, and
one can try to think of these degrees of freedom in terms of free
quasi-particles. The average separation between the quasi-particles is
equal to
\be \label{avsep}
\Delta X \sim \frac{\ell}{r_H}\, l_{p(d)}
\ee
with $l_{p(d)}$ the $d$-dimensional Planck length. If the
quasi-particles move with the speed of light and $\sigma$ represents the
probability that quasi-particles will interact with the endpoint of the
string, then we expect a mean free path time of the order of \be
\label{aux9} t_{\rm mfp} \sim {\Delta X\over \sigma}.  \ee

Supposing for a moment we assume that this is the same as the mean free
path time on the boundary which, in Appendix \ref{sec:t_mfp}, we argued
to be given by $t_{\rm mfp}\sim 1/(T\sqrt{\lambda})$.  Combined with
(\ref{aux9}) and (\ref{avsep}) this leads to the interaction probability
\be \label{intprob}
\sigma  \sim \frac{d-1}{4\pi}\, \frac{\ell}{l_s^2}\, l_{d(p)}
\ee
which for the usual AdS${}_5$ case leads to a scaling with $g_s$
and $N$ as $\sigma \sim g_s^{1/2} N^{-1/6}$. This is a rather
peculiar prediction for the interaction strength of the string
endpoint with the quasi-particles. Since the quasi-particles are
made out of strings (or branes) some $g_s$ dependence is to be
expected, and the interaction strength indeed vanishes as $g_s
\rightarrow 0$.

In deriving \eqref{intprob} we have assumed that the mean free path time
on the stretched horizon is identical to that on the boundary. This
however, is unlikely to pertain as we explain now. In fact, we will
argue that on general grounds we should expect that $t_{\rm mfp}\sim
1/T$ on any stretched horizon. The logic relies on using dimensional
analysis coupled with thermal physics of black holes. Generically we
expect,
\be \label{stmfp}
t_{\rm mfp} = \frac{1}{T}\,\CG
\!\left(Tl_s, {l_{p(d)}\over l_s}, {\ell\over l_s}\right) 
\ee
where $\CG$ is a function of the dimensionless ratios of the length
scales available. We have fixed the overall normalization to be
determined by the thermal scale on physical grounds. Furthermore, using
the facts that: (i) the dynamics of string probes generically are
unaware of the Planck scale (to determine which we could for instance
use D-brane probes \cite{Douglas:1996yp}) and, (ii) the geometry near a
black hole horizon is the Rindler spacetime, which is insensitive to the
cosmological constant, we can argue that $\CG \sim 1$, {\it i.e.}, it is
independent of the hierarchy between the various length scales in the
problem.  More precisely, in the near horizon region, ${r-r_H\over
r_H}\ll 1$, the AdS$_d$ black hole metric \eqref{AdSdBH} reduces to the
Rindler metric:
\begin{align}
 ds_d^2\approx - \tilde r^2 d\tilde t^2+d\tilde r^2
 +d\vec{\tilde X}_{d-2}^2,\label{NHAdSd}
\end{align}
where
\begin{align}
 \tilde t=2\pi T t, \qquad
 \tilde r=\sqrt{r-r_H\over \pi T},\qquad
 \vec{\tilde X}_{d-2}={r_H\over\ell}\vec{X}_{d-2}.
\end{align}
Because the metric \eqref{NHAdSd} does not contain any scale such as
$\ell$, the dynamics of a fundamental string in the near horizon region
can only depend on $l_s$ (dependence on $l_{p(d)}$ is excluded as in
(i)).  Therefore, the mean free path time $\tilde t_{\rm mfp}$
determined from the dynamics of the string can only depend on $l_s$.
However, because $\tilde t_{\rm mfp}$ is dimensionless, it should be
that $\tilde t_{\rm mfp}\sim 1$, which means $t_{\rm mfp}\sim 1/T$.  One
can give a more concrete argument by using the argument in Appendix
\ref{sec:t_mfp} applied to the near-horizon geometry \eqref{NHAdSd}.

Now using $t_{\rm mfp} \sim 1/T$ we can conclude that the interaction
probability only depends on the ratio of the $d$-dimensional Planck
scale and string scale:
\be \label{intprob2}
\sigma  \sim \frac{d-1}{4\pi}\, \frac{l_{p(d)}}{l_s}\ ,
\ee
which suggests a universal dynamics of the stretched horizon independent
of the asymptotics of the spacetime. Nevertheless \eqref{intprob2} leads
to an interaction probability which is a non-trivial function of $g_s$
as $l_{p(d)}$ depends non-trivially on the details of the
compactification.\footnote{ Actually, there is no consensus on where to
place the stretched horizon.  For example, Refs.\ \cite{Iizuka:2002wa,
Iizuka:2003ad} explained some thermodynamical properties of black holes
by postulating the existence of quasi-particles living on a stretched
horizon a distance $l_{p(d)}$ away from the horizon, instead of $l_s$.
In this case with a stretched horizon located $l_{p(d)}$ away from the
horizon, we obtain a simpler result $\sigma\sim 1$ instead of
\eqref{intprob2}.  This simple form of $\sigma$ is appealing, but we do
not know of a physical reason to choose one over the other. }

Clearly, it would be interesting to explore this line of thought
further and, for example, also find an interpretation for the
collision time. However, many of the assumptions we made are
highly questionable. For example, we ignored backreaction, and
only used the quadratic part of the Nambu--Goto action. The latter
approximation certainly breaks down once we are a proper distance
$\sim l_s$ away from the horizon. It is also unclear to what
extent we can really think of the stretched horizon as a gas of
almost free quasi-particles. We leave an exploration of these
issues to future work.

\section{Discussion}
\label{sec:disc}

In this paper, we discussed Brownian motion in the holographic
context, in order to shed light on near-equilibrium dynamics of
strongly coupled thermal gauge theories. A useful probe exhibiting
Brownian motion consists of a fundamental string stretching
between the boundary and the horizon and being randomly excited by
the black hole environment.  We established the relation between
the observables associated with such Brownian particle in the
boundary theory and those of the transverse mode excitations of
the fundamental string.  At the semiclassical level, the modes on
the string are thermally excited due to Hawking radiation and,
consequently, the motion of the boundary Brownian particle is
described by a Langevin equation, which involves stochastic force
and friction.  In the bulk, the stochastic force corresponds to
the random excitation of the string by the Hawking radiation,
while the friction corresponds to the fact that the excitations on
the string get dissipated into the horizon.

Although in this paper we focused on the relation at the
semiclassical level between the boundary Brownian motion and the
dynamics of the fundamental string in the bulk, the boundary-bulk
dictionary we wrote down in subsection \ref{ss:bdybulk} in
principle allows one to predict the precise correlations of the
Hawking radiation quanta beyond the semiclassical approximation,
in terms of the precise correlation functions for the boundary
Brownian particle. Obtaining the latter of course requires one to
compute correlation functions in strongly coupled plasmas, which
is a difficult task. Nevertheless, such a dictionary is an
important step toward understanding the microphysics underlying
the fluid-gravity correspondence.

One of the particularly interesting results of the current paper
is the estimate in subsection \ref{sec:HolAuto} for the time
scales associated with the Brownian particle immersed in a CFT
plasma:
\begin{align}
 t_{\rm relax}\sim {m\over \sqrt{\lambda}\,T^2},\qquad
 t_{\rm coll}\sim {1\over T},\qquad
 t_{\rm mfp}\sim {1\over \sqrt{\lambda}\,T}.
\end{align}
Note that setting $m=T$ in $t_{\rm relax}$ gives $t_{\rm mfp}$, which is
a consistency check because a fluid particle can be thought of as a
Brownian particle with mass $\sim T$.  The fact that $t_{\rm coll}\gg
t_{\rm mfp}$ at strong coupling $\lambda\gg 1$ implies that a Brownian
particle interacts with many plasma particles simultaneously.  Because
of this, a Brownian particle with mass $m\ll \sqrt{\lambda}\,T$ can
thermalize in a time much shorter than $t_{\rm coll}$, the time elapsed
in a single process of collision.
This is reminiscent of the recent conjecture
\cite{Hayden:2007cs,Sekino:2008he} that black holes can scramble
information very fast, whose dual picture is that a degree of freedom in
the boundary theory interacts with a huge number of other degrees
of freedom simultaneously.  It would be interesting to study this possible
connection further.

Historically, the main achievement of the theory of Brownian
motion was the determination of the value of the Avogadro constant
$\CN_A=6\times 10^{23}\,{\rm mol}^{-1}$, which is huge but finite.
If $\CN_A$ were infinite, the diffusion constant would be zero and
we would not be able to observe Brownian motion.  The fact that we
can observe it in nature gives evidence that $\CN_A$ is finite and
fluids are not continuous but made of molecules.
Then, what is the analogue of the Avogadro constant in the
Brownian motion in the AdS/CFT context we studied, and what is the
bulk significance of it?
In the case of AdS$_5$/SYM$_4$, the macroscopic energy density of
the plasma scales as $E=\CO(N^2)$, while the energy carried by a
microscopic quantum is of the order of the temperature
$T=\CO(N^0)$. What corresponds to the Avogadro constant is the
ratio of these, $N^2/N^0=N^2$.  The finiteness of $\CN_A$
corresponds to the finiteness of $N$.
In the bulk, on the other hand, what corresponds to $E$ is the
mass of the black hole, $M\sim R_s/G_N\sim \CO(G_N^{-1})$ with
$R_s$ the Schwarzschild radius, while $T$ is the Hawking
temperature $T_H\sim \CO(G_N^0)$.  The ratio is
$M/T_H=\CO(G_N^{-1})=\CO(N^2)$.  So, in the bulk, the finiteness
of $\CN_A$ corresponds to the finiteness of $G_N$, or to the fact
that the energy carried by a Hawking radiation quantum is finite
although it is much smaller than the mass of the black hole.

We considered non-relativistic Brownian motion in the current
paper, which is the result of the quadratic approximation we made
in \eqref{SNG} to the Nambu--Goto action.  It would be interesting
to generalize our treatment to the relativistic case, where
Brownian motion and its Langevin dynamics are not very well
understood; for a recent discussion, see {\it e.g.}\
\cite{Dunkel:2008}.
Such a generalization can also be regarded as a generalization of
the drag force computations of \cite{Herzog:2006gh,
Gubser:2006bz}, which are relativistic because the full
Nambu--Goto action was taken into account, to non-stationary
($\omega\neq 0$) solutions.
Also, with such a relativistic formalism, one can presumably give a more
rigorous derivation of $t_{\rm mfp}$ than the one we did in Appendix
\ref{sec:t_mfp}.

As explained above, the stochastic force appearing in the Langevin
equation is related to the friction term via the
fluctuation-dissipation theorem.  In the bulk the latter is
mimicked by the dissipative nature of the event horizon, which is
present for all black holes.  On the other hand, the stochastic
term arises due to the Hawking temperature of the black hole; yet
only non-extremal black holes have finite Bekenstein--Hawking
temperature.  This leads to the naive puzzle that whereas the
dissipation is always present, fluctuation is seemingly absent for
extremal black holes since these have zero temperature. It would
be interesting to see whether the quantum fluctuations which are
present even at zero temperature suffice to account for the origin
of the stochastic processes. Note that this is not a-priori
unreasonable in the AdS/CFT context; although quantum processes
are $1/N$ suppressed in the large $N$ field theory, we had to
account for the semi-classical Hawking radiation phenomena to see
the origin of the random force in the Langevin equation.
Furthermore, extremal black holes could also be subject to
super-radiant type instabilities which can effectively mimic the
physics of Hawking radiation. In fact, this feature has been
exploited recently to show how the microstate `geometries' can
reproduce some features of the thermal Hawking spectrum
\cite{Chowdhury:2008uj}.

The stochastic random force which drives the long time diffusive
motion has a characteristic dependence on the temperature, which
we derived assuming that the system was thermodynamically stable.
As is well known, considering the global as opposed to the
Poincar\'e patch of AdS provides two distinct black hole solutions
at the same temperature---the small black hole which has negative
specific heat and a large black hole which is in thermal
equilibrium with the Hawking radiation. To be able to access both
these solutions simultaneously one has to work in the global AdS
geometry which has a compact spatial boundary. The physics of a
probe string endpoint in the small black hole background should
exhibit marked differences from the Brownian motion discussed
above, despite the system experiencing the same temperature.  In a
finite volume system we naively expect the Brownian process to
saturate after the time scale $t = \frac{\pi^3 \, \ell^4\, T
}{\ap}$, for in this time the particle has diffused throughout the
system.  In the bulk this presumably corresponds to the string
diffusing out completely on the stretched horizon and becoming
indistinguishable from the thermal atmosphere. This can in fact be
used to probe the difference between the large and the small black
hole. Imagine we normalize the physics of the string endpoint on
the boundary to correspond to the Brownian motion undertaken in
the large black hole. Using this as the UV boundary condition for
the probe string in the small black hole background, we can
examine the dynamics of the endpoint at the IR stretched horizon.
A plausible conjecture for this dynamics is that the fluctuations
of the string are macroscopically large on the stretched horizon,
in fact will have a scale comparable to the black hole itself.

%
%

\section*{Acknowledgments}

We would like to thank V.~Balasubramanian, J.~Casalderrey-Solana,
I.~Kanitscheider, E.~Keski-Vakkuri, P.~Kraus, N.~Iizuka, T.~Levi,
P.~McFadden, T.~McLoughlin, S.~Minwalla, A.~Paredes, A.~Parnachev,
K.~Peeters, E.~Verlinde, M.~Zamaklar, and especially K.~Papadodimas for
valuable discussions.
We would also like to thank the organizers of the workshops
``Gravitational Thermodynamics and the Quantum Nature of Space''
at the University of Edinburgh, and ``Black Holes: A Landscape of
Theoretical Physics Problems'' at CERN, for stimulating
environments.
V.H. and M.R. are supported in part by STFC\@. The work of M.S.
was supported by an NWO Spinoza grant. The work of J.d.B. and M.S.
is partially supported by the FOM foundation.

\appendix

\section{Normalized basis}
\label{sec:norm_basis}

In this appendix, we discuss the quantization of the action
\eqref{SNG} obtained from the Nambu--Goto action and derive the
normalized basis of solutions \eqref{uomega_norm'd} to the
equation of motion.

\subsection{Canonical commutation relations and normalized basis}

The canonical commutation relations for the theory \eqref{SNG}:
\begin{align}
  S_{\text{NG}}^{(2)}=-{1\over 4\pi\ap}\int d^2 x\sqrt{-g(x)}\,
 g^{\mu\nu}(x)G_{IJ}(x)
 {\partial X^I\over \partial x^\mu}{\partial X^J\over \partial x^\nu},\qquad
 x^\mu=t,r,
\end{align}
are given by
\begin{equation}
\begin{split}
 & [X^I(x),X^J(x')]_\Sigma=0,\qquad
 [X^I(x),n^\mu \partial_\mu X^J(x')]_\Sigma=i{2\pi\ap\over \sqrt{h}}G^{IJ}\delta(r-r'),\\
 &[n^\mu \partial_\mu X^I(x),n^\nu \partial_\nu X^J(x')]_\Sigma=0.
\end{split}
\label{CCR}
\end{equation}
Here, $\Sigma$ is a Cauchy surface in the $x^\mu=t,r$ part of the
spacetime \eqref{genmetric}, $h_{ij}$ is the metric on $\Sigma$
induced from $g_{\mu\nu}$, and $n^\mu$ is the future-pointing unit
normal to $\Sigma$.
For functions $f^I(x),g^I(x)$ satisfying the equation of motion
\eqref{eom_NG}, we can define the following inner product:
\begin{align}
 (f,g)_\Sigma&=-{i\over 2\pi\ap}\int_\Sigma dx\sqrt{h}\, n^\mu
 G_{IJ}(f^I  \partial_\mu g^{J*} - \partial_\mu f^I \,g^{J*}).
\label{innerprod}
\end{align}
It can be shown that this inner product is independent of the
choice of $\Sigma$, just as the standard Klein--Gordon inner
product \cite{Birrell:1982ix}.  This inner product satisfies
\begin{gather}
  (f,g)^*=-(f^*,g^*)=(g,f),\\
 (af_1+bf_2,g)^*=a(f_1,g)+b(f_2,g),\quad
 (f,ag_1+bg_2)^*=a^*(f,g_1)+b^*(f,g_2).
\end{gather}
It is not difficult to show that the canonical commutation
relations \eqref{CCR} are equivalent to
\begin{align}
 [(f,X)_\Sigma,(g,X)_\Sigma]_\Sigma=(f,g^*)_\Sigma\quad
 \text{$\forall f,g$ satisfying the equation of motion \eqref{eom_NG}}.\label{CCRcond_innerprod}
\end{align}
Let $\{u_\alpha^I(x)\}$ be a basis of normalized functions
satisfying the equation of motion \eqref{eom_NG} such that
\begin{align}
 (u_\alpha,u_\beta)=
 -(u_\alpha^*,u_\beta^*)=\delta_{\alpha\beta},\qquad
 (u_\alpha,u_\beta^*)=0,
\end{align}
and expand $X^I$ as
\begin{align}
 X^I(x)=\sum_\alpha \left[a_\alpha u_\alpha^I(x)+a_\alpha^\dagger u_\alpha^I(x)^*\right].\label{X^Iexpninualpha}
\end{align}
Then one can readily show that the condition
\eqref{CCRcond_innerprod} implies
\begin{align}
 [a_\alpha,a_\beta]=[a_\alpha^\dagger,a_\beta^\dagger]=0,\qquad
 [a_\alpha,a_\beta^\dagger]=\delta_{\alpha\beta}.
\end{align}

\subsection{Normalized basis for AdS$_3$}

As shown in the main text, in the AdS$_3$ case, the solution to
the equation of motion can be written as (see Eq.\
\eqref{A[fp+Bfm]})
\begin{align}
 u_\omega(t,\rho)&=
 A\Bigl[f^{(+)}_\omega(\rho)+Bf^{(-)}_\omega(\rho)\Bigr] e^{-i\omega t},
\end{align}
where $B$ satisfies boundary conditions \eqref{BoverA} at
$\rho=\rho_c$ and \eqref{BoverAnh} at $\rho=1+2\epsilon$.   The
inner product \eqref{innerprod} for this solution is
\begin{align}
 (u_\omega,u_\omega)
 &={2\omega \ell^2 |A|^2\over \ap\beta}
 \left[ {2\rho\over 1+\rho^2\nu^2}+\log\left({\rho-1\over \rho+1}\right)\right]^{\rho=\rho_c}_{\rho=1+2\epsilon}
 \approx
 {2\omega \ell^2 |A|^2\over \ap\beta}
 \log\left({1\over \epsilon}\right).
\end{align}
From this, one obtains the normalized basis:
\begin{align}
 u_\omega(t,\rho)=
 \sqrt{\ap\beta\over 2\ell^2\omega\log(1/\epsilon)}
 \Bigl[f^{(+)}_\omega(\rho)+{B}f^{(-)}_\omega(\rho)\Bigr] e^{-i\omega t}.
\end{align}

\section{Evaluation of displacement squared $s^2_{\rm reg}(t)$}
\label{sec:eval_s^2}

In this appendix, we evaluate the displacement squared
\eqref{s_reg^2}, which can be written as:
\begin{align}
  s^2_{\rm reg}(t)&=
 {4\ap\beta^2 \over \pi^2 \ell^2}
 \int_0^\infty {d\nu\over\nu}\,{1+\nu^2\over 1+\rho_c^2\nu^2}
  {\sin^2{\pi t\nu\over \beta}\over e^{2\pi \nu}-1}
 =
 {\ap\beta^2 \over \pi^2 \ell^2}
\left( {\rho_c^2-1\over \rho_c^2} I_1 +{1\over \rho_c^2}  I_2
 \right),
 \label{s_reg^2_app}
\end{align}
where
\begin{align}
 \begin{split}
 I_1
 &= 4 \int_0^\infty {dx \over x(1+a^2x^2)}  {\sin^2{kx\over 2}\over e^{x}-1}
 = \int_{-\infty}^\infty {dx \over |x|(1+a^2x^2)}  {1-e^{ikx}\over e^{|x|}-1},\\
 I_2
 &= 4 \int_0^\infty {dx \over x}  {\sin^2{kx\over 2}\over e^{x}-1}
 = \int_{-\infty}^\infty {dx \over |x|}  {1-e^{ikx}\over e^{|x|}-1},
 \end{split}
 \label{I1I2}
\end{align}
and we defined new variables by
\begin{align}
 x=2\pi \nu,\qquad a={\rho_c\over 2\pi},\qquad k={t\over \beta}.
\end{align}

The integrals \eqref{I1I2} can be evaluated using the standard
method of deforming the contour on the complex $x$ plane.  For
that, one first replaces $|x|$ with $\sqrt{x^2+\epsilon^2}$ with
$\epsilon$ a small positive number.  If $k>0$, one can then deform
the contour to run vertically around the branch cut between
$i\epsilon$ and $\infty$.  The resulting integral is simpler than
\eqref{I1I2} and, after taking $\epsilon\to 0$, can be
analytically evaluated.  One should also take into account the
contribution from the poles of the integrand on the imaginary
axis.
The final result is
\begin{align}
 \begin{split}
 I_1
 & =
 {1\over 2}\left[e^{k/ a}{\rm Ei}(-{k\over a})+e^{-k/a}{\rm Ei}({k\over a})\right]
 +{1\over 2}\left[\psi(1+{1\over 2\pi a})+\psi(1-{1\over 2\pi a})\right]
 \\
 &\qquad
 +{e^{-2\pi |k|} \over 2}\left[
 {{}_2F_1(1,1+{1\over 2\pi a}; 2+{1\over 2\pi a}; e^{-2\pi |k|}) \over 1+{1\over 2\pi a}}
 +
 {{}_2F_1(1,1-{1\over 2\pi a}; 2-{1\over 2\pi a}; e^{-2\pi |k|}) \over 1-{1\over 2\pi a}}
 \right]
\\
 &\qquad
 -{\pi\over 2}(1-e^{-|k|/a})\cot{1\over 2a}
 +\log\left({2a \sinh\pi k\over k}\right),
 \\
 I_2
 &=\log\left({\sinh \pi k\over \pi k}\right).
 \end{split}
 \label{I1I2_expl}
\end{align}
where ${\rm Ei}(z)$ is the exponential integral,
${}_2F_1(\alpha,\beta;\gamma;z)$ is the hypergeometric function,
and $\psi(z)=(d/ dz)\log\Gamma(z)$ is the digamma function.  For
${\rm Ei}(z)$, we take a branch where both ${\rm Ei}(x>0)$ and
${\rm Ei}(x<0)$ are real.

If $\rho_c\gg 1$ and thus $a\gg 1$, one can use the expressions
\eqref{I1I2_expl} to derive the following behavior:
\begin{align}
 I_1 =
 \begin{cases}
  {\pi k^2\over 2a} + \CO(a^{-2})\\
  \pi k+\CO(\log k)
 \end{cases}
 \qquad
 I_2 =
 \begin{cases}
  \CO(a^{0})        & \qquad(k\ll a)\\
  \pi k+\CO(\log k) & \qquad(k\gg a)
 \end{cases}
\end{align}
Therefore, if $\rho_c\gg 1$, $s_{\rm reg}^2(t)$ has the following
behavior:
\begin{align}
 s_{\rm reg}^2(t) =
 \begin{cases}
  {\ap \over \ell^2\rho_c}t^2    + \CO({1\over \rho_c^{2}}) & \qquad(t\ll \beta)\\[1ex]
  {\ap \beta \over \pi \ell^2 }t + \CO(\log{t\over \beta})  & \qquad(t\gg\beta)
 \end{cases}
\end{align}

\section{Distribution of momentum $p$}
\label{sec:p-distribution}

In this appendix, we compute the probability distribution of the
momentum $p=m\dot x$, where $x$ is the position of the string endpoint
at the UV cut-off $\rho=\rho_c$, and show that it is exactly equal to
the Maxwell--Boltzmann distribution.\footnote{Here, we will ignore the
fact that the mass of the quark gets corrected in thermal medium
\cite{Herzog:2006gh}, and make a crude estimate by using the bare mass
\eqref{m_and_rhoc}.  }

From \eqref{X_rhoc}, the momentum of the particle is
\begin{align}
 p&=m \dot x(t)
 =-{im\over \ell}\sum_{\omega>0}
 \sqrt{2\ap\beta\omega \over \log(1/\epsilon)}
 \left[
 {1-i\nu \over 1-i\rho_c \nu}\left({\rho_c-1\over \rho_c+1}\right)^{i\nu/2}e^{-i\omega t}a_\omega
 -{\rm h.c.}
 \right]
\end{align}
We would like to know the probability distribution $f(p)$ of $p$.
By definition,
\begin{align}
 \ev{e^{ip\xi}}=\int_{-\infty}^\infty dp\, e^{ip\xi} f(p).
\end{align}
Namely, $f(p)$ is the Fourier transform of $\ev{e^{ip\xi}}$.  So,
what we want to compute is
\begin{align}
 \ev{{:\!e^{ip\xi}\!:}}
 &=
 \Ev{:\!
  \exp \left\{
 {\xi m\over \ell}
 \sum_{\omega>0}
 \sqrt{2\ap\beta\omega \over \log(1/\epsilon)}
 \left[
 {1-i\nu \over 1-i\rho_c \nu}\left({\rho_c-1\over \rho_c+1}\right)^{i\nu/2}e^{-i\omega t}a_\omega
 -{\rm h.c.}
 \right]\right\}
 \!:},
\end{align}
where we regularized the operator by normal ordering.  The
expectation value is with respect to the density matrix
\eqref{rho0}.  Using the identity
\begin{align}
 \tr\left[e^{-\beta\omega a^\dagger a}\, {:\!e^{\alpha a-\alpha^* a^\dagger}\!\!:}\,\right]
 ={1\over 1-e^{-\beta\omega}}\,
 {\exp\bigl(-{|\alpha|^2\over e^{\beta\omega}-1}\bigr)},
\end{align}
we can compute
\begin{align}
 \ev{{:\!e^{ip\xi}\!:}}
 &=C
 \exp \left[
 -{2\xi^2\ap m^2\over \ell^2}
 \int_0^\infty \!\! {d\nu\,\nu(1+\nu^2) \over (1+\rho_c^2 \nu^2)(e^{2\pi\nu}-1)}
 \right]
\label{febw17Oct08},
\end{align}
where $C$ is a constant independent of $\xi$ and we rewrote the
sum over $\omega$ by an integral using \eqref{Deltaomega}. This
integral can be evaluated by deforming the contour in the complex
plane, just as we did for \eqref{I1I2}, the result being
\begin{align}
 \int_0^\infty \!\! {d\nu\,\nu(1+\nu^2) \over (1+\rho_c^2 \nu^2)(e^{2\pi\nu}-1)}
 ={\rho_c^2-1\over 4\rho_c^4}\left[
 \pi \cot{\pi\over \rho_c}-\psi(1+{1\over \rho_c})-\psi(1-{1\over \rho_c})-2\log \rho_c
 \right]
 +{1\over 24\rho_c^2},
\end{align}
where $\psi(z)=(d/dz)\log\Gamma(z)$ is the digamma function.
Using this expression, it is easy to show that, for large
$\rho_c$,
\begin{align}
 \int_0^\infty {d\nu\, \nu (1+\nu^2) \over (1+\rho_c^2 \nu^2)(e^{2\pi\nu}-1)}
 ={1\over 4\rho_c}+\dots.
\end{align}
Therefore, from \eqref{febw17Oct08}, we obtain
\begin{align}
 \ev{:\!e^{ip\xi}\!:}
 &=C e^{-m \xi^2 / 2\beta}
\end{align}
for large $\rho_c$, where we used \eqref{m_and_rhoc}.  By Fourier
transforming,
\begin{align}
 f(p)&= \int_{-\infty}^\infty {d\xi\over 2\pi} e^{-ip\xi} \ev{:\!e^{ip\xi}\!:}
 \propto e^{-\beta E_p},\qquad
 E_p\equiv {p^2\over 2m}.
\end{align}
This is exactly the Maxwell--Boltzmann distribution of particles
with energy $E_p$.  Therefore, for large $\rho_c$, the endpoint of
the string behaves like a non-relativistic particle with mass $m$
immersed in a thermal bath of temperature $T$.

\section{Mean free path time $t_{\rm mfp}$}
\label{sec:t_mfp}

In subsection \ref{sec:HolAuto}, we discussed the time scales associated
with Brownian motion: the relaxation time $t_{\rm relax}$ and the
collision duration time $t_{\rm coll}$.  In this Appendix, we evaluate
$t_{\rm mfp}$, the mean free path time, or the typical time between two
collisions, using the correlators of the random force $R(t)$ in the case of AdS$_3$.
We argue that the mean free path
time is given by
\begin{align}
 t_{\rm mfp}\sim {1\over \sqrt{\lambda}\,T},
\end{align}
where $\lambda\sim \ell^4/\ap^2$ is the 't Hooft coupling, although we are unable to give
a rigorous derivation.  We expect that this holds in more general cases,
including AdS$_5$.

In subsection \ref{ss:t_mfp:gen}, we discuss how to determine
characteristic time scales from correlators in general.  In subsection
\ref{ss:t_mft:comp}, we compute $t_{\text{mfp}}$ for the Brownian motion
in the case of AdS$_3$.

\subsection{Correlators and time scales}
\label{ss:t_mfp:gen}

\begin{figure}[htbp]
  \begin{quote}
 \begin{center}
  \epsfxsize=6cm \epsfbox{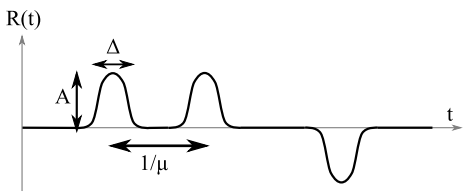} \caption{\sl A sample of the
 stochastic variable $R(t)$, which consists of many pulses randomly
 distributed.}  \label{fig:pulses}
\end{center}\end{quote}
\end{figure}

Consider a stochastic quantity $R(t)$ whose functional form consists of
many pulses randomly distributed.  Let the form of a single pulse be
$f(t)$, with width $\Delta$ and amplitude $A$\@.  Furthermore, assume
that the pulses come with random signs.  If we have $k$ pulses at
$t=t_i$ ($i=1,2,\dots,k$), then $R(t)$ is given by
\begin{align}
 R(t)&=\sum_{i=1}^k \epsilon_i f(t-t_i),
\end{align}
where $\epsilon_i=\pm 1$ are random signs.  For a schematic picture, see
Figure \ref{fig:pulses}.
Let us assume that the distribution of pulses obeys the Poisson
distribution.  Namely, the probability that there are $k$ pulses in an
interval of length $\tau$, say $[0,\tau]$, is given by
\begin{align}
 P_k(\tau)=e^{-\mu\tau}{(\mu\tau)^k\over k!}.
\end{align}
Here, $\mu$ is the number of pulses per unit time.  In other words,
$1/\mu$ is the average distance between two pulses.  We do not assume
that the pulses are well separated; namely, we do \emph{not} assume
$\Delta\ll 1/\mu$.  Later, we will identify $R(t)$ with the random force
in the Langevin equation; the pulses are contributions from a collision
with a fluid particle, and therefore $t_{\rm mfp}=1/\mu$.

The 2-point function for $R$ can be written as
\begin{align}
 \ev{R(t)R(t')}=\sum_{k=1}^\infty 
 e^{-\mu\tau}{(\mu\tau)^k\over k!}
 \sum_{i,j=1}^k \ev{\epsilon_i \epsilon_j f(t-t_i)f(t'-t_j)}_k,\label{bjse22Dec08}
\end{align}
where we assumed $t,t'\in[0,\tau]$ and $\ev{~}_k$ is the statistical
average when there are $k$ pulses during $[0,\tau]$.  Because $k$ pulses
are randomly and independently distributed in the interval $[0,\tau]$,
this expectation value is computed as
\begin{multline}
 \sum_{i,j=1}^k \ev{\epsilon_i \epsilon_j f(t-t_i)f(t'-t_j)}_k
\\
 =
 {1\over \tau^k}
 \int_0^\tau dt_1\cdots dt_k
 \left[
 \sum_{i=1}^k f(t-t_i)f(t'-t_i)
 +\sum_{i\neq j}^k \ev{\epsilon_i \epsilon_j}_k f(t-t_i)f(t'-t_j)
 \right].
\end{multline}
Here, the second term vanishes because $\ev{\epsilon_i\epsilon_j}_k=0$
for $i\neq j$.  Therefore, one readily computes
\begin{align}
 \sum_{i,j=1}\ev{\epsilon_i \epsilon_j f(t-t_i)f(t'-t_j)}_k
 &=
 {k\over \tau}
 \int_0^\tau dt_1 f(t-t_1)f(t'-t_1)\notag\\
 &\approx
 {k\over \tau}
 \int_{-\infty}^\infty  dt_1\, f(t-t'-t_1)f(-t_1)
 \equiv {k\over \tau} F(t-t').
\end{align}
Here, in going to the second line, we took $\tau$ to be much larger 
than the width $\Delta$ of $f(t)$, which is always possible because
$\tau$ is arbitrary. Substituting this back into \eqref{bjse22Dec08}, we
find
\begin{align}
 \ev{R(t)R(t')}=\mu F(t-t').
\end{align}

In a similar way, one can compute the following 4-point function:
\begin{align}
 \ev{R^2(t)R^2(t')}=\sum_{k=1}^\infty 
 e^{-\mu\tau}{(\mu\tau)^k\over k!}
 \sum_{i,j,m,n=1}^k 
 \ev{\epsilon_i \epsilon_j \epsilon_m \epsilon_n
 f(t-t_i)f(t-t_j)f(t'-t_m)f(t'-t_n)}_k.\label{blqk22Dec08}
\end{align}
Again, the expectation value $\ev{\epsilon_i \epsilon_j \epsilon_m
\epsilon_n}_k$ vanishes unless some of $i,j,m,n$ are equal.  The
possibilities are $i=j\neq m=n$, $i=m\neq j=n$, $i=n\neq j=m$, and
$i=j=m=n$.  Therefore,
\begin{align}
 &\sum_{i,j,m,n=1}^k 
 \ev{\epsilon_i \epsilon_j \epsilon_m \epsilon_n
 f(t-t_i)f(t-t_j)f(t'-t_m)f(t'-t_n)}_k
 \notag\\
 &=
 \Bigl\langle
 \sum_{i\neq j}^k \left[f(t-t_i)^2 f(t'-t_j)^2
 +2f(t-t_i)f(t'-t_i)f(t-t_j)f(t'-t_j)\right]
 \notag\\[-2ex]
 &\qquad\qquad
 +\sum_{i=1}^k f(t-t_i)^2f(t'-t_i)^2
 \Bigr\rangle_k
 \notag\\
 &=
 {k(k-1)\over \tau^2}
 \int_{-\infty}^\infty  dt_1dt_2 \left[f(t-t_1)^2 f(t'-t_2)^2
 + f(t-t_1)f(t'-t_1)f(t-t_2)f(t'-t_2)\right]
 \notag\\
 &\qquad\qquad
 +{k\over \tau} \int_{-\infty}^\infty dt_1 f(t-t_1)^2f(t'-t_1)^2.
\end{align}
Substituting this back into \eqref{blqk22Dec08}, we obtain
\begin{align}
 \ev{R^2(t)R^2(t')}
 &=
 \mu^2 [F(0)^2 +F(t-t')^2]+
 \mu  \int_{-\infty}^\infty du \, f(t-t'-u)^2f(-u)^2\notag\\
 &=
 \ev{R^2(t)}\ev{R^2(t')}+\ev{R(t)R(t')}^2
 +\mu  \int_{-\infty}^\infty du \, f(t-t'-u)^2f(-u)^2.
\end{align}

For example, consider the following shape function for the pulse
\begin{align}
 f(t)=A e^{-t^2/2\Delta^2}.\label{gauss_f(t)}
\end{align}
Then, one computes
\begin{align}
 \ev{R(t)R(0)}&=\sqrt{\pi}\,\Delta\,\mu\,  A^2 e^{-t^2/4\Delta^2},\label{bmjn22Dec08}\\
 \ev{R^2(t)R^2(0)}&=
 \ev{R^2(t)}\ev{R^2(0)} +\ev{R(t)R(0)}^2
 +{1\over \sqrt{2\pi}\,\Delta\,\mu}\ev{R(t)R(0)}^2.\label{bskz22Dec08}
\end{align}
Therefore, if we know the behavior of $\ev{R(t)R(0)}$, we can read off
$\Delta$ and $\mu A$ from \eqref{bmjn22Dec08}.  If we further know
$\ev{R^2(t)R^2(0)} $ then, from \eqref{bskz22Dec08}, we can read off
$\mu$. In particular, if we denote the last term in \eqref{bskz22Dec08}
by $\ev{R^4}'$, then
\begin{align}
 \mu^{-1}\sim {\ev{R^4}' \over \ev{R^2}^2}\,\Delta.
 \label{readoffmu}
\end{align}
This result \eqref{readoffmu} is expected to be true for other forms of
$f(t)$, not just for the Gaussian case \eqref{gauss_f(t)}.

Note that the treatment above is classical.  If $R(t)$ is a quantum
operator, we should consider the classical part of the correlators by
appropriately subtracting quantum divergences.

\subsection{Evaluation of $t_{\rm mfp}$ for Brownian motion}
\label{ss:t_mft:comp}

Let us evaluate $t_{\rm mfp}$ for the boundary Brownian motion, by
identifying the stochastic function $R(t)$ in the previous subsection
with the random force appearing in the Langevin equation.  A pulse
$f(t)$ corresponds to the contribution from a collision with a single
plasma particle.  $\Delta$ is the time elapsed in a single collision,
namely $\Delta=t_{\rm coll}$, while $1/\mu$ is the time between two
collisions, namely $1/\mu=t_{\rm mfp}$.

Using Eq.\ \eqref{X_rhoc} as well as the relations $p=m\dot{x}$ and
$R(\omega)=p(\omega)/\mu(\omega)$, where $\mu(\omega)$ is given by
\eqref{boundary_mu}, we can write the random force $R(t)$ as
\begin{align}
 R(t)&=\sum_{\omega>0}(r_\omega\, e^{i\omega t}\,a_\omega+\text{h.c.}),\qquad
 r_\omega 
 =
 -i\sqrt{8\pi^2\ell^2\omega\over \ap\beta^3\log(1/\epsilon)}\,
 {1-i\nu\over 1-i\nu/\rho_c}\left({\rho_c-1\over\rho_c+1}\right)^{\!\! {i\nu\over 2}}.\label{R_expn}
\end{align}
Because $a_{\omega}$ are free harmonic oscillators, it is easy to show
that
\begin{align}
 \ev{:\! R^2(t)R^2(0)\!:}
 &=
 \ev{:\!R^2(t)\!:}\ev{:\!R^2(0)\!:} +\ev{:\!R(t)R(0)\!:}^2.\label{buyh22Dec08}
\end{align}
Here, we are considering the normal-ordered correlators because the
result of the previous subsection applies to the classical piece of
correlators; henceforth, normal ordering of operators will be
understood. By comparing \eqref{buyh22Dec08} with \eqref{bskz22Dec08},
we appear to have $t_{\rm mfp}=0$.
However, this is due to the non-relativistic approximation we made in
\eqref{SNG} when we expanded the Nambu--Goto action up to quadratic
order.  If we keep the next order (quartic) terms, we obtain the
following additional contribution to the Hamiltonian:
\begin{align}
 H^{(4)}=-{1\over 16\pi \ap}\int_{r_s}^{r_c} dr
 \left[{(\partial_t X)^2\over h(r)}-{r^4h(r)\over \ell^4}(\partial_r X)^2\right]^2,\qquad
 h(r)=1-\left({r_H\over r}\right)^2,
\end{align}
where $r_s=(1+2\epsilon)\, r_H$, $\epsilon\ll 1$.  This corresponds to
the first relativistic correction to the non-relativistic action
$S^{(2)}_{\text{NG}}$.  In the presence of this interaction, there is an
extra contribution to the correlator $\ev{R^2(t)R^2(0)}$ coming from the
contractions with the terms in $H^{(4)}$.  If we consider the case with
$t=0$, we have
\begin{align}
 \ev{R^2(0)R^2(0)}\equiv \ev{ R^4}
 =2\ev{R^2}_0^2
 -\beta\ev{R^4 H^{(4)}}_0,
 \label{ncxs28Dec08}
\end{align}
where $\ev{~}_0$ is the expectation value with respect to the quadratic
action $S^{(2)}_{\text{NG}}$, {\it i.e.}, it is the expectation value
with respect to the density matrix \eqref{rho0}.

So, let us evaluate the last term in \eqref{ncxs28Dec08}, which will be
denoted by $\ev{R^4}'$.  Using the expansions \eqref{X_expn} and
\eqref{R_expn}, the explicit expression for $\ev{R^4}'$ is
\begin{align}
 \ev{R^4}'
 &=
 -{\beta\over 16\pi \ap}
 \Biggl\langle
 \sum_{\omega_1,\dots,\omega_4>0}
 (r_{\omega_1} a_{\omega_1}+{\rm h.c.})
 (r_{\omega_2} a_{\omega_2}+{\rm h.c.})
 (r_{\omega_3} a_{\omega_3}+{\rm h.c.})
 (r_{\omega_4} a_{\omega_4}+{\rm h.c.})\notag\\
 & \times
\int_{r_s}^{r_c} dr \left\{
 {1\over h}\Bigl[\sum_{\omega>0} \omega(u_\omega a_\omega-u_\omega^* a_\omega^\dagger)\Bigr]^2
 +{r^4h\over \ell^4} 
 \Bigl[\sum_{\omega>0} \bigl((\partial_r u_\omega) a_\omega+(\partial_r u_\omega^*) a_\omega^\dagger\bigr)\Bigr]^2
 \right\}^2
 \Biggr\rangle_0.
\end{align}
There are many terms coming from the expansion of this. Let us focus on
the following term in particular:
\begin{align}
 &-{\beta\over 16\pi\ap}
 \sum_{\omega_1,\dots,\omega_4} \sum_{\omega_1',\dots,\omega_4'}
 \omega_1\, \omega_2\, \omega_3\, \omega_4 \,
 r_{\omega_1}^* r_{\omega_2}^* r_{\omega_3}^* r_{\omega_4}
 \ev{
   a_{\omega_1}^\dagger a_{\omega_2}^\dagger a_{\omega_3}^\dagger a_{\omega_4}
   a_{\omega_1'} a_{\omega_2'} a_{\omega_3'} a_{\omega_4'}^\dagger
 }_0
 \int_{r_s}^{r_c} {dr \over h^2}\,
 u_{\omega_1'}
 u_{\omega_2'}
 u_{\omega_3'}
 u_{\omega_4'}^*
 .\label{mqsw29Dec08}
\end{align}
There are various ways to contract $a,a^\dagger$.  Let us take the term
obtained by contracting $a_{\omega_i}$ against $a_{\omega_i'}^\dagger$,
or $a_{\omega_i}^\dagger$ against $a_{\omega_i'}$, where $i=1,\dots,4$.
Other contractions give similar contributions. This particular
contraction gives the following:
\begin{align}
 &-{\beta\over 16\pi\ap}
 \sum_{\omega_1,\dots,\omega_4}
 \omega_1\, \omega_2\, \omega_3\, \omega_4 \,
 r_{\omega_1}^* r_{\omega_2}^* r_{\omega_3}^* r_{\omega_4}
 \left[\prod_{i=1}^4{1\over e^{\beta \omega_i}-1}\right]
 \int_{r_s}^{r_c} {dr \over h^2}\,
 u_{\omega_1}
 u_{\omega_2}
 u_{\omega_3}
 u_{\omega_4}^*
 \notag
 \\
 &\qquad\qquad
 \sim 
 {1\over \ap\beta^3}
 \sum_{\omega_1,\dots,\omega_4\lesssim \beta^{-1}}
 r_{\omega_1}^* r_{\omega_2}^* r_{\omega_3}^* r_{\omega_4}
 \int_{r_s}^{r_c} {dr \over h^2}\,
 u_{\omega_1} u_{\omega_2} u_{\omega_3} u_{\omega_4}^*,\label{jyit29Dec08}
\end{align}
where the Bose--Einstein factor $1/(e^{\beta \omega_i}-1)$ has
effectively cut off the $\omega_i$ sum at $\beta^{-1}$.  From now on, we
do not keep track of numerical factors.

We would like to evaluate the $r$ integral in \eqref{jyit29Dec08}.
Because the integrand in \eqref{jyit29Dec08} has a second order pole at
$r=r_H$ due to $h^{-2}$, the dominant contribution comes from $r\sim
r_s\approx r_H$.
For a while, let us instead consider the case where there is \emph{a
first order pole at $r=r_H$, by replacing $h^{-2}$ by $h^{-1}$}.  From
\eqref{X_near_hor} and \eqref{tortoise}, near $r=r_H$,
\begin{align}
 u_\omega
 &\approx
 \sqrt{\ap\beta\over 2\ell^2\omega \log(1/\epsilon)}
 (e^{i\omega r_*}+e^{i\theta_\omega}e^{-i\omega r_*})e^{-i\omega t},\qquad
 {dr\over h}\approx {r_H^2\over \ell^2}dr_*={4\pi^2 \ell^2\over \beta^2}dr_*.
\end{align}
Therefore, the integral in \eqref{jyit29Dec08} (with $h^{-2}$ replaced
by $h^{-1}$) is
\begin{align}
 &\sim {\ap^2 \over\ell^2 [\log(1/\epsilon)]^2\sqrt{\omega_1\omega_2\omega_3\omega_4}}
 \int _{-{\beta\over 4\pi}\log({1\over \epsilon})} dr_*\,
 [e^{ i\omega_1 r_*}+e^{ i\theta_{\omega_1}}e^{-i\omega_1 r_*}]
 [e^{ i\omega_2 r_*}+e^{ i\theta_{\omega_2}}e^{-i\omega_2 r_*}]\notag\\
 &\qquad\qquad
 \times
 [e^{ i\omega_3 r_*}+e^{ i\theta_{\omega_3}}e^{-i\omega_3 r_*}]
 [e^{-i\omega_4 r_*}+e^{-i\theta_{\omega_4}}e^{ i\omega_4 r_*}]\,
 e^{-i(\omega_1+\omega_2+\omega_3-\omega_4) t}\label{glax29Dec08}
\end{align}
The dominant part in the $\epsilon\to 0$ limit can be easily evaluated
by noting that
\begin{align}
 \int _{-{\beta\over 4\pi}\log({1\over\epsilon})} dr_*\, e^{i\omega r_*} =
 \delta_{\omega,0}\,{\beta\over 4\pi}\log\Bigl({1\over \epsilon}\Bigr) 
 +\text{(finite as $\epsilon\to 0$)}.
 \label{glas29Dec08}
\end{align}
For example, by collecting the first terms in the four pairs of the
brackets in \eqref{glax29Dec08}, one finds
\begin{align}
 &\sim {\ap^2\beta \over\ell^2 \log(1/\epsilon)\,\sqrt{\omega_1\omega_2\omega_3\omega_4}}
 \,
 \delta_{\omega_1+\omega_2+\omega_3,\,\omega_4}.
\end{align}
Note that the finite part in \eqref{glas29Dec08} does not survive in the
$\epsilon\to 0$ limit.  If we plug this result back into
\eqref{jyit29Dec08}, using the explicit expression for $r_\omega$ in
\eqref{R_expn}, we find
\begin{align}
 \sim {\ell^2\over \ap\beta^8[\log(1/\epsilon)]^3}\sum_{\omega_1,\omega_2,\omega_3\lesssim\beta^{-1}}
 \biggl[{1-i\nu_1\over 1-i \nu_1/\rho_c}\left({\rho_c-1\over \rho_c+1}\right)^{i\nu_1/2}\biggr]
 \biggl[~2~\biggr]
 \biggl[~3~\biggr]
 \biggl[~1+2+3~\biggr]^*.
\end{align}
Here, ``$[\,2\,]$'' denotes the previous factor with $\nu_1$ replaced by
$\nu_2$.  ``$[\,3\,]$'' and ``$[\,1+2+3\,]$'' are similar.  By rewriting
the sum by integral using \eqref{Deltaomega} and using the fact that
$\rho_c\gg 1$, this is estimated as
\begin{align}
 \sim {\ell^2\over \ap\beta^{5}}\int_{\lesssim\beta^{-1}}d\omega_1\,d\omega_2\,d\omega_3
 \sim {\ell^2\over \ap\beta^{8}}\sim {1\over \beta^8\sqrt{\lambda}},\label{R4estimate1}
\end{align}
where we used the relation $\lambda\sim\ell^4/\ap^2$. There are many
other terms we did not discuss, such as other contractions of
\eqref{mqsw29Dec08}, but these will not affect this estimate.

However, of course, this is not precisely what we wanted to evaluate; we
have replaced $h^{-2}$ in \eqref{jyit29Dec08} by $h^{-1}$.  However,
using $h^{-2}$ instead will change the above discussion, because the $r$
integral around $r=r_H$ will now give a power ($\sim 1/\epsilon$)
divergence instead of the logarithmic divergence we had in
\eqref{glas29Dec08}.  This log divergence was important in obtaining the
result \eqref{R4estimate1}, because this log divergence was canceled
against the normalization factor in $u_\omega\sim
[\log(1/\epsilon)]^{-1/2}$.  What we have forgotten is that, if we
include the quartic correction $H^{(4)}$, we should also consider
corrections to the normalized basis $u_\omega$, which presumably
introduces a normalization factor that goes as $\epsilon^{1/2}$.  This
corrected normalization factor should cancel against the power
divergence coming from $h^{-2}$, thus giving a finite result, which
should give \eqref{R4estimate1} at the end of the day---namely,
\begin{align}
  \ev{R^4}'\sim{1\over \beta^8\sqrt{\lambda}}.
 \label{R4estimate2}
\end{align}
Whatever the modifications due to the quartic term are, the dominant
contribution comes from the region $r\sim r_H$ and quantities such as
$r_c$ or $m$ cannot enter the final result.  Also, the relativistic
correction must come with a factor of $\ap\sim \lambda^{-1/2}$.  There
being no other available quantities, $\ev{R^4}'$ must be proportional to
\eqref{R4estimate2}.
A fully relativistic formalism in which one can rigorously and
explicitly show \eqref{R4estimate2} is beyond the scope of the current
paper.  We leave development of such a formalism for future research.

From \eqref{kappa^n}, we have 
\begin{align}
 \ev{R^2}=\kappa^{\rm n}(t=0)\sim {\ell^2\over \ap\beta^4}
\sim {1\over \beta^4\sqrt{\lambda}}.
\end{align}
Therefore, using the formula \eqref{readoffmu} with $\Delta=\beta$, we obtain
\begin{align}
 t_{\rm mfp}
 \sim {\beta\over \sqrt{\lambda}}
 ={1\over T\sqrt{\lambda}}.
 \label{tmfp_est}
\end{align}
It is satisfactory that this does not depend on the properties of the
Brownian particle probe such as $m$, because $t_{\rm mfp}$ is a time
scale associated with the fluid itself.

\section{Solving equation of motion for general $d$ using matching technique}
\label{app:gen_dim}

In this appendix, we solve the wave equation \eqref{fundeq2} for
general dimensions using the matching technique for low
frequencies $\omega\ll T$.  We would like to obtain a solution
which is purely ingoing at the horizon and, in particular,
determine its behavior near the boundary.

Let us denote by $X^{-}_{\omega}(r)$ this particular solution of
the wave equation \eqref{fundeq2} which obeys the purely ingoing
boundary condition at the horizon $r=r_H$. To determine it, let us
consider three regions: (A) a near horizon region with $r\sim r_H$
and $V(r)\ll \omega^2$, (B) an intermediate region with $V(r)\gg
\omega^2$, and (C) an asymptotic region with $r\gg r_H$.
The idea is to consider the approximate solutions in each of the
three regions, and to match these to each other. For more details,
see \cite{Harmark:2007jy} and references therein.
As before, we define
\be \rho \equiv \frac{r}{r_H},\qquad \nu\equiv \frac{\ell^2\,  \omega}{r_H}.
\ee
In terms of these parameters the constraints on the different
regions under consideration, $V(r)\ll \omega^2$ and $V(r)\gg
\omega^2$ respectively translate to $\rho-1\ll\nu^2$ and
$\rho-1\gg\nu^2$.  Furthermore, the low frequency condition,
$\omega\ll T$, can be written as $\nu\ll 1$.

In region A, where $\rho-1\ll\nu^2$ and we can drop the potential
$V(r)$ from \eqref{fundeq2}, the linearly independent solutions
are
\begin{align}
X_A^\pm(r) = e^{\pm \,i\,\omega\, r_*} \sim \exp\left[\pm
{i\,\nu\over d-1}\log(\rho-1) \right]. \label{sol_A-B}
\end{align}
The purely ingoing solution is $X^-_A(r) = e^{-i\,\omega \,
r_{\ast}}$. Now, since $\nu \ll 1$ we can focus on a region
slightly away from the horizon (still remaining in region A), such
that $\exp(-{\nu\over d-1})\ll \rho-1 \ll \nu^2$.  Here we can
approximate the purely ingoing solution as
\be
X^-_A(\rho) \sim 1 - \frac{i\,\nu}{d-1} \, \log(\rho-1) .
 \label{sol_regionA}
\ee

In the asymptotic region C, where $\rho\gg 1$, we can approximate
$h(r)\sim 1$. The linearly independent solutions of (\ref{fundeq})
are then
\be
X^{\pm}_C(\rho) = \left(1\mp \frac{i\nu}{\rho}\right) e^{\pm
i\,\nu/\rho}. \label{indepsol_C2}
\ee
The general solution can be written as
\begin{align}
 X_C=C^+\, X^+_C + C^- \, X^-_C  \ ,
\end{align}
which, for $\rho\gg 1$, can be expanded as
\begin{align}
 X_C
 &=
 (C^+ + C^-)\left(1+\frac{\nu^2}{2\rho^2} + \ldots\right)
 +(C^+ - C^-)\left(\frac{i}{3} \frac{\nu^3}{\rho^3}+\dots\right).\label{regionC-B}
\end{align}

Finally, in region B, where $\rho-1\gg \nu^2$ and we can drop
$\omega^2$ from \eqref{fundeq2}, leading then to the general
solution
\be
X_{B}(\rho) = B_1 + B_2 \int^{\rho}_{\infty} \frac{d\rho'}{\rho'^4
h(\rho')}, \label{regionB}
\ee
where $B_1$ and $B_2$ are two integration constants.  For
$\rho\sim 1$ (but still $\rho-1\gg \nu^2$), we can approximate
$h(\rho)\sim (d-1)(\rho-1)$ and \eqref{regionB} gives
\begin{align}
 X_B(\rho)= B_1+B_2\left[{1\over d-1}\log(\rho-1)+b\right],
\label{regionB-A}
\end{align}
where $b$ is a constant independent of $\nu$ whose precise value
is not relevant for our purpose.

We now have the solutions in the three regions A--C; by matching
them across the domains of overlap we can relate the various
constants of integration.  To begin with we determine $B_1$ and
$B_2$ by matching \eqref{regionB-A} in region B with the solution
\eqref{sol_regionA} in region A, obtaining
\be B_1=1+i\, b\, \nu, \qquad B_2 = -\,i\, \nu.
\label{sol_B}
\ee
To determine $C^\pm$ we expend the solution in region B
\eqref{regionB} for $\rho \gg 1$ and match it to that in region C
\eqref{regionC-B} leading to
\begin{align}
 B_1=C^+ + C^-,\qquad
 B_2=-{i\nu^3}(C^+ - C^-).
\label{sol_B2}
\end{align}
It must be borne in mind that we have performed the matching only
in the small frequency limit $\nu \ll1$ and as a result should
trust the expressions only at the leading order in $\nu$.
Solving \eqref{sol_B} and \eqref{sol_B2}, we finally find that the
purely ingoing solution behaves at large $\rho$ as
\be
X^-_\omega(\rho)  = C^+\, X^+_C + C^- \,X^-_C,\qquad C^{\pm} =
\frac{1}{2}\,\left(1\pm\frac{1}{\nu^2} +i\,b\,\nu\right).
\label{X-omega2}
\ee



\begin{thebibliography}{99}

\bibitem{Brown}
R.~Brown, ``A brief account of microscopical observations made in
the months of June, July and August, 1827, on the particles
contained in the pollen of plants; and on the general existence of
active molecules in organic and inorganic bodies,'' Philos.\
Mag.\, {\bf 4}, 161 (1828); reprinted in Edinburgh New Philos.\
J.\ {\bf 5}, 358 (1928).

\bibitem{Uhlenbeck:1930zz}
  G.~E.~Uhlenbeck and L.~S.~Ornstein,
  ``On The Theory Of The Brownian Motion,''
  Phys.\ Rev.\  {\bf 36}, 823 (1930).

\bibitem{Chandrasekhar:1943ws}
  S.~Chandrasekhar,
  ``Stochastic problems in physics and astronomy,''
  Rev.\ Mod.\ Phys.\  {\bf 15}, 1 (1943).

\bibitem{UhlenbeckII}
M.~C.~Wang and G. E. Uhlenbeck, ``On the Theory of the Brownian
Motion II,'' Rev.\ Mod.\ Phys.\ {\bf 17}, 323 (1945).

\bibitem{Dunkel:2008}
J.~Dunkel and P.~H\"anggi, ``Relativistic Brownian Motion,''
arXiv:0812.1996 [cond-mat].

\bibitem{Kappler:1931}
E.~Kappler, ``Versuche zur Messung der Avogadro-Loschmidtschen
Zahl aus der Brownschen Bewegung einer Drehwaage,'' Ann.\ Phys.\
(Leipzig), {\bf
    403},
233 (1931).

\bibitem{Maldacena:1997re}
  J.~M.~Maldacena,
  ``The large N limit of superconformal field theories and supergravity,''
  Adv.\ Theor.\ Math.\ Phys.\  {\bf 2} (1998) 231
  [Int.\ J.\ Theor.\ Phys.\  {\bf 38} (1999) 1113]
  [arXiv:hep-th/9711200].

\bibitem{Gubser:1998bc}
  S.~S.~Gubser, I.~R.~Klebanov and A.~M.~Polyakov,
  ``Gauge theory correlators from non-critical string theory,''
  Phys.\ Lett.\  B {\bf 428}, 105 (1998)
  [arXiv:hep-th/9802109].

\bibitem{Witten:1998qj}
  E.~Witten,
  ``Anti-de Sitter space and holography,''
  Adv.\ Theor.\ Math.\ Phys.\  {\bf 2}, 253 (1998)
  [arXiv:hep-th/9802150].

\bibitem{Aharony:1999ti}
  O.~Aharony, S.~S.~Gubser, J.~M.~Maldacena, H.~Ooguri and Y.~Oz,
  ``Large N field theories, string theory and gravity,''
  Phys.\ Rept.\  {\bf 323} (2000) 183
  [arXiv:hep-th/9905111].

\bibitem{Bhattacharyya:2008jc}
  S.~Bhattacharyya, V.~E.~Hubeny, S.~Minwalla and M.~Rangamani,
  ``Nonlinear Fluid Dynamics from Gravity,''
  JHEP {\bf 0802}, 045 (2008)
  [arXiv:0712.2456 [hep-th]].

\bibitem{Son:2007vk}
  D.~T.~Son and A.~O.~Starinets,
  ``Viscosity, Black Holes, and Quantum Field Theory,''
  Ann.\ Rev.\ Nucl.\ Part.\ Sci.\  {\bf 57}, 95 (2007)
  [arXiv:0704.0240 [hep-th]].


\bibitem{Strominger:1996sh}
  A.~Strominger and C.~Vafa,
  ``Microscopic Origin of the Bekenstein-Hawking Entropy,''
  Phys.\ Lett.\  B {\bf 379}, 99 (1996)
  [arXiv:hep-th/9601029].

\bibitem{Maldacena:1997de}
  J.~M.~Maldacena, A.~Strominger and E.~Witten,
  ``Black hole entropy in M-theory,''
  JHEP {\bf 9712}, 002 (1997)
  [arXiv:hep-th/9711053].

\bibitem{Mathur:2005zp}
  S.~D.~Mathur,
 ``The fuzzball proposal for black holes: An elementary review,''
  Fortsch.\ Phys.\  {\bf 53}, 793 (2005)
  [arXiv:hep-th/0502050].

\bibitem{Bena:2007kg}
  I.~Bena and N.~P.~Warner,
  ``Black holes, black rings and their microstates,''
  Lect.\ Notes Phys.\  {\bf 755}, 1 (2008)
  [arXiv:hep-th/0701216].

\bibitem{Skenderis:2008qn}
  K.~Skenderis and M.~Taylor,
  ``The fuzzball proposal for black holes,''
  Phys.\ Rept.\  {\bf 467}, 117 (2008)
  [arXiv:0804.0552 [hep-th]].

\bibitem{Balasubramanian:2008da}
  V.~Balasubramanian, J.~de Boer, S.~El-Showk and I.~Messamah,
 ``Black Holes as Effective Geometries,''
  Class.\ Quant.\ Grav.\  {\bf 25}, 214004 (2008)
  [arXiv:0811.0263 [hep-th]].

\bibitem{Kovtun:2004de}
  P.~Kovtun, D.~T.~Son and A.~O.~Starinets,
  ``Viscosity in strongly interacting quantum field theories from black hole
  physics,''
  Phys.\ Rev.\ Lett.\  {\bf 94}, 111601 (2005)
  [arXiv:hep-th/0405231].


\bibitem{Herzog:2006gh}
  C.~P.~Herzog, A.~Karch, P.~Kovtun, C.~Kozcaz and L.~G.~Yaffe,
  ``Energy loss of a heavy quark moving through N = 4 supersymmetric
  Yang-Mills plasma,''
  JHEP {\bf 0607}, 013 (2006)
  [arXiv:hep-th/0605158].

\bibitem{Liu:2006ug}
  H.~Liu, K.~Rajagopal and U.~A.~Wiedemann,
  ``Calculating the jet quenching parameter from AdS/CFT,''
  Phys.\ Rev.\ Lett.\  {\bf 97}, 182301 (2006)
  [arXiv:hep-ph/0605178].

\bibitem{Gubser:2006bz}
  S.~S.~Gubser,
  ``Drag force in AdS/CFT,''
  Phys.\ Rev.\  D {\bf 74}, 126005 (2006)
  [arXiv:hep-th/0605182].


\bibitem{Herzog:2006se}
  C.~P.~Herzog,
  ``Energy loss of heavy quarks from asymptotically AdS geometries,''
  JHEP {\bf 0609}, 032 (2006)
  [arXiv:hep-th/0605191].

\bibitem{CasalderreySolana:2006rq}
  J.~Casalderrey-Solana and D.~Teaney,
  ``Heavy quark diffusion in strongly coupled N = 4 Yang Mills,''
  Phys.\ Rev.\  D {\bf 74}, 085012 (2006)
  [arXiv:hep-ph/0605199].


\bibitem{Gubser:2006nz}
  S.~S.~Gubser,
  ``Momentum fluctuations of heavy quarks in the gauge-string duality,''
  Nucl.\ Phys.\  B {\bf 790}, 175 (2008)
  [arXiv:hep-th/0612143].

\bibitem{Liu:2006he}
  H.~Liu, K.~Rajagopal and U.~A.~Wiedemann,
  ``Wilson loops in heavy ion collisions and their calculation in AdS/CFT,''
  JHEP {\bf 0703}, 066 (2007)
  [arXiv:hep-ph/0612168].

\bibitem{CasalderreySolana:2007qw}
  J.~Casalderrey-Solana and D.~Teaney,
  ``Transverse momentum broadening of a fast quark in a N = 4 Yang Mills
  plasma,''
  JHEP {\bf 0704}, 039 (2007)
  [arXiv:hep-th/0701123].

\bibitem{RHICreviews}
  D.~Mateos,
  ``String Theory and Quantum Chromodynamics,''
  Class.\ Quant.\ Grav.\  {\bf 24}, S713 (2007)
  [arXiv:0709.1523 [hep-th]]. \\
  S.~S.~Gubser,
  ``Heavy ion collisions and black hole dynamics,''
  Gen.\ Rel.\ Grav.\  {\bf 39}, 1533 (2007)
  [Int.\ J.\ Mod.\ Phys.\  D {\bf 17}, 673 (2008)].\\
  D.~T.~Son,
  ``Gauge-gravity duality and heavy-ion collisions,''
  AIP Conf.\ Proc.\  {\bf 957} (2007) 134.\\
  J.~D.~Edelstein and C.~A.~Salgado,
  ``Jet Quenching in Heavy Ion Collisions from AdS/CFT,''
  AIP Conf.\ Proc.\  {\bf 1031}, 207 (2008)
  [arXiv:0805.4515 [hep-th]].

\bibitem{Moore:2004tg}
  G.~D.~Moore and D.~Teaney,
  ``How much do heavy quarks thermalize in a heavy ion collision?,''
  Phys.\ Rev.\  C {\bf 71}, 064904 (2005)
  [arXiv:hep-ph/0412346].

\bibitem{Myers:2007we}
  R.~C.~Myers, A.~O.~Starinets and R.~M.~Thomson,
``Holographic spectral functions and diffusion constants for fundamental matter,''
  JHEP {\bf 0711}, 091 (2007)
  [arXiv:0706.0162 [hep-th]].


\bibitem{Rey:1998bq}
  S.~J.~Rey, S.~Theisen and J.~T.~Yee,
 ``Wilson-Polyakov loop at finite temperature in large N gauge theory and
  anti-de Sitter supergravity,''
  Nucl.\ Phys.\  B {\bf 527}, 171 (1998)
  [arXiv:hep-th/9803135].

\bibitem{Kubo:f-d_thm}
R.~Kubo, ``The fluctuation-dissipation theorem,'' Rep.\ Prog.\
Phys.\ {\bf 29}, 255-284 (1966).

\bibitem{Mori:genLE}
H.~Mori, ``Transport, collective motion, and Brownian motion,''
Prog.\ Theor.\ Phys.\ {\bf 33}, 423 (1965).

\bibitem{Balasubramanian:1998sn}
  V.~Balasubramanian, P.~Kraus and A.~E.~Lawrence,
  ``Bulk vs. boundary dynamics in anti-de Sitter spacetime,''
  Phys.\ Rev.\  D {\bf 59}, 046003 (1999)
  [arXiv:hep-th/9805171].

\bibitem{Hemming:2000as}
  S.~Hemming and E.~Keski-Vakkuri,
  ``Hawking radiation from AdS black holes,''
  Phys.\ Rev.\  D {\bf 64}, 044006 (2001)
  [arXiv:gr-qc/0005115].

\bibitem{Lawrence:1993sg}
  A.~E.~Lawrence and E.~J.~Martinec,
  ``Black Hole Evaporation Along Macroscopic Strings,''
  Phys.\ Rev.\  D {\bf 50}, 2680 (1994)
  [arXiv:hep-th/9312127].

\bibitem{Frolov:2000kx}
  V.~P.~Frolov and D.~Fursaev,
  ``Mining energy from a black hole by strings,''
  Phys.\ Rev.\  D {\bf 63}, 124010 (2001)
  [arXiv:hep-th/0012260].

\bibitem{Birrell:1982ix}
  N.~D.~Birrell and P.~C.~W.~Davies,
  ``Quantum Fields In Curved Space,''
{\it  Cambridge, UK: Univ.\ Pr.\ (1982) 340p}

\bibitem{Karliner:1988hd}
  M.~Karliner, I.~R.~Klebanov and L.~Susskind,
  Int.\ J.\ Mod.\ Phys.\  A {\bf 3}, 1981 (1988).

\bibitem{KTH}
R.~Kubo, M.~Toda, and N.~Hashitsume, ``Statistical Physics II --
    Nonequilibrium Statistical Mechanics,'' Springer-Verlag.

\bibitem{Thorne:1986iy}
  K.~S.~Thorne, R.~H.~Price and D.~A.~Macdonald,
  ``Black Holes: The Membrane Paradigm,''
{\it  New Haven, USA: Yale Univ. Pr. (1986) 367p}

\bibitem{Kovtun:2003wp}
  P.~Kovtun, D.~T.~Son and A.~O.~Starinets,
  ``Holography and hydrodynamics: Diffusion on stretched horizons,''
  JHEP {\bf 0310}, 064 (2003)
  [arXiv:hep-th/0309213].

\bibitem{Saremi:2007dn}
  O.~Saremi,
  ``Shear waves, sound waves on a shimmering horizon,''
  arXiv:hep-th/0703170.

\bibitem{Fujita:2007fg}
  M.~Fujita,
  ``Non-equilibrium thermodynamics near the horizon and holography,''
  JHEP {\bf 0810}, 031 (2008)
  [arXiv:0712.2289 [hep-th]].

\bibitem{Starinets:2008fb}
  A.~O.~Starinets,
 ``Quasinormal spectrum and the black hole membrane paradigm,''
  arXiv:0806.3797 [hep-th].


\bibitem{Iqbal:2008by}
  N.~Iqbal and H.~Liu,
  ``Universality of the hydrodynamic limit in AdS/CFT and the membrane paradigm,''
  arXiv:0809.3808 [hep-th].

\bibitem{Parikh:1997ma}
  M.~Parikh and F.~Wilczek,
  ``An action for black hole membranes,''
  Phys.\ Rev.\  D {\bf 58}, 064011 (1998)
  [arXiv:gr-qc/9712077].

\bibitem{Eling:2006aw}
  C.~Eling, R.~Guedens and T.~Jacobson,
  ``Non-equilibrium Thermodynamics of Spacetime,''
  Phys.\ Rev.\ Lett.\  {\bf 96}, 121301 (2006)
  [arXiv:gr-qc/0602001].


\bibitem{Susskind:1993ws}
  L.~Susskind,
  ``Some speculations about black hole entropy in string theory,''
  arXiv:hep-th/9309145.

\bibitem{Halyo:1996vi}
  E.~Halyo, A.~Rajaraman and L.~Susskind,
  ``Braneless black holes,''
  Phys.\ Lett.\  B {\bf 392}, 319 (1997)
  [arXiv:hep-th/9605112].

\bibitem{Horowitz:1996nw}
  G.~T.~Horowitz and J.~Polchinski,
  ``A correspondence principle for black holes and strings,''
  Phys.\ Rev.\  D {\bf 55}, 6189 (1997)
  [arXiv:hep-th/9612146].


\bibitem{Douglas:1996yp}
  M.~R.~Douglas, D.~N.~Kabat, P.~Pouliot and S.~H.~Shenker,
  ``D-branes and short distances in string theory,''
  Nucl.\ Phys.\  B {\bf 485}, 85 (1997)
  [arXiv:hep-th/9608024].


\bibitem{Iizuka:2002wa}
  N.~Iizuka, D.~N.~Kabat, G.~Lifschytz and D.~A.~Lowe,
  ``Quasiparticle picture of black holes and the entropy-area relation,''
  Phys.\ Rev.\  D {\bf 67}, 124001 (2003)
  [arXiv:hep-th/0212246].

\bibitem{Iizuka:2003ad}
  N.~Iizuka, D.~N.~Kabat, G.~Lifschytz and D.~A.~Lowe,
  ``Stretched horizons, quasiparticles and quasinormal modes,''
  Phys.\ Rev.\  D {\bf 68}, 084021 (2003)
  [arXiv:hep-th/0306209].

\bibitem{Hayden:2007cs}
  P.~Hayden and J.~Preskill,
  ``Black holes as mirrors: quantum information in random subsystems,''
  JHEP {\bf 0709}, 120 (2007)
  [arXiv:0708.4025 [hep-th]].


\bibitem{Sekino:2008he}
  Y.~Sekino and L.~Susskind,
  ``Fast Scramblers,''
  JHEP {\bf 0810}, 065 (2008)
  [arXiv:0808.2096 [hep-th]].


\bibitem{Chowdhury:2008uj}
  B.~D.~Chowdhury and S.~D.~Mathur,
``Non-extremal fuzzballs and ergoregion emission,''
  arXiv:0810.2951 [hep-th].

\bibitem{Harmark:2007jy}
  T.~Harmark, J.~Natario and R.~Schiappa,
  ``Greybody Factors for d-Dimensional Black Holes,''
  arXiv:0708.0017 [hep-th].


\end{thebibliography}
\end{document}